\documentclass[11pt, a4paper]{article}

\usepackage{jcappub_ver}
\usepackage{enumerate}
\usepackage{amsbsy}
\usepackage{graphics}
\usepackage{mathrsfs}
\usepackage{wrapfig}
\newcommand{\be}{\begin{equation}} \newcommand{\ee}{\end{equation}}
\newcommand{\bea}{\begin{eqnarray}} \newcommand{\eea}{\end{eqnarray}}
\newcommand{\el}{\nonumber \\}
\newcommand{\re}[1]{(\ref{#1})}

\newcommand{\pat}{\partial}

\renewcommand{\sec}[1]{section \ref{#1}}
\newcommand{\fig}[1]{figure \ref{#1}}
\newcommand{\brt}[1]{[#1]}
\newcommand{\para}{\paragraph}

\renewcommand{\a}{\alpha}
\renewcommand{\b}{\beta}
\renewcommand{\c}{\gamma}
\renewcommand{\d}{\delta}
\newcommand{\e}{\epsilon}
\renewcommand{\l}{\lambda}

\newcommand{\LCDM}{$\Lambda$CDM\ }
\newcommand{\GN}{G_{\mathrm{N}}}

\newcommand{\rmd}{\mathrm{d}}
\newcommand{\diag}{\mathrm{diag}}
\newcommand{\bz}{\bar{z}}

\newcommand{\rb}{r_{\mathrm{b}}}
\newcommand{\tB}{{t_{\mathrm{B}}}}
\newcommand{\Em}{{|E|_{\mathrm{max}}}}
\newcommand{\sign}{\mathrm{sign}}
\newcommand{\Rb}{R_{\mathrm{b}}}

\newcommand{\nonum}{\\}
\newcommand{\etal}{et al.\ }
\newcommand{\ie}{i.e.\ }

\newcommand{\adot}{\dot{a}}

\newcommand{\Rdot}{\dot{R}}

\newcommand{\cp}{c_{\phi}}

\newcommand{\Hfit}{H_\mathrm{fit}}
\newcommand{\wHtot}{w_{H\mathrm{tot}}}
\newcommand{\wDtot}{w_{D\mathrm{tot}}}
\newcommand{\ba}{\bar{a}}
\newcommand{\bH}{\bar{H}}
\newcommand{\badot}{\dot{\bar{a}}}
\newcommand{\baddot}{\ddot{\bar{a}}}

\newcommand{\thetat}{\tilde{\theta}}
\newcommand{\sigmat}{\tilde{\sigma}}
\newcommand{\htt}{\tilde{h}}
\newcommand{\patl}[1]{\frac{\rmd{#1}}{\rmd\l}}

\newcommand{\zb}{z_\mathrm{b}}

\newcommand{\av}[1]{\langle{#1}\rangle}
\newcommand{\sQ}{\mathcal{Q}}
\newcommand{\sR}{{^{(3)}R}}

\newcommand{\Om}{\Omega_{\mathrm{m}}}
\newcommand{\Omn}{\Omega_{\mathrm{m0}}}
\newcommand{\Omfit}{\Omega_{\mathrm{mfit}}}

\newcommand{\OQ}{\Omega_{\sQ}}
\newcommand{\OQn}{\Omega_{\sQ0}}
\newcommand{\OR}{\Omega_{R}}
\newcommand{\ORn}{\Omega_{R0}}

\newcommand{\sO}{\mathcal{O}}

\newcommand{\mC}{\mathscr{C}}
\newcommand{\mE}{\mathscr{E}}
\newcommand{\rmax}{r_\mathrm{max}}
\newcommand{\rmin}{r_\mathrm{min}}
\newcommand{\DA}{D_\mathrm{A}}

\newcommand{\rhom}{\rho_{\mathrm{m}}}

\newcommand{\PRD}[1]{{\it Phys. Rev.} {\bf D#1}}

\newcommand{\PRL}[1]{{\it Phys. Rev. Lett.} {\bf #1}}

\newcommand{\PLA}[1]{{\it Phys. Lett.} {\bf A#1}}
\newcommand{\PLB}[1]{{\it Phys. Lett.} {\bf B#1}}
\newcommand{\MNRAS}[1]{{\it Mon. Not. Roy. Astron. Soc.} {\bf #1}}
\newcommand{\APJ}[1]{{\it Astrophys. J.} {\bf #1}}
\newcommand{\APJS}[1]{{\it Astrophys. J. Suppl.} {\bf #1}}

\newcommand{\CQG}[1]{{\it Class. Quant. Grav.} {\bf #1}}
\newcommand{\GRG}[1]{{\it Gen. Rel. Grav.} {\bf #1}}
\newcommand{\AaA}[1]{{\it Astron. \& Astrophys.} {\bf #1}}
\newcommand{\PROG}[1]{{\it Prog. Theor. Phys.} {\bf #1}}

\newcommand{\IJMPD}[1]{{\it Int. J. Mod. Phys.} {\bf D#1}}

\title{Average expansion rate and light propagation in a cosmological Tardis spacetime}

\author[a]{Mikko Lavinto,}

\author[a]{Syksy R\"{a}s\"{a}nen,}

\author[b]{and Sebastian J. Szybka}

\affiliation[a]{University of Helsinki, Department of Physics \\
and Helsinki Institute of Physics \\
P.O. Box 64, FIN-00014 University of Helsinki, Finland}

\affiliation[b]{Astronomical Observatory, Jagellonian University \\
Orla 171, 30-244 Krak\'ow, Poland}

\emailAdd{mikko {\it dot} lavinto {\it at} helsinki {\it dot} fi}
\emailAdd{syksy {\it dot} rasanen {\it at} iki {\it dot} fi}
\emailAdd{sebastian {\it dot} szybka {\it at} uj {\it dot} edu {\it dot} pl}

\abstract{We construct the first exact statistically homogeneous
and isotropic cosmological solution in which inhomogeneity has
a significant effect on the expansion rate.
The universe is modelled as a Swiss Cheese, with
dust FRW background and inhomogeneous holes.
We show that if the holes are described by the
quasispherical Szekeres solution, their average
expansion rate is close to the background under certain
rather general conditions.
We specialise to spherically symmetric holes and
violate one of these conditions.
As a result, the average expansion rate at late times
grows relative to the background, \ie backreaction is significant.
The holes fit smoothly into the background, but are larger
on the inside than a corresponding background domain:
we call them {\em Tardis regions}.
We study light propagation, find the effective
equations of state and consider the relation
of the spatially averaged expansion rate to the
redshift and the angular diameter distance.
}

\begin{document}

\maketitle
  
\setcounter{tocdepth}{2}

\setcounter{secnumdepth}{3}

\section{Introduction} \label{sec:intro}

\para{The backreaction conjecture.}

Predictions of homogeneous and isotropic models of the universe
with ordinary matter (with non-negative pressure) and ordinary
gravity (based on the four-dimensional Einstein-Hilbert
action) disagree with observations of cosmological distances and
average expansion rate at late times by a factor of two 
(see \cite{Buchert:2011} for discussion and references).
The problem is usually addressed by introducing exotic matter with
negative pressure or modifying gravity on large scales, leading
to accelerated expansion.
However, it is possible that the failure of the predictions of
homogeneous and isotropic models is related to the known
breakdown of homogeneity and isotropy due to structure
formation, rather than unknown fundamental physics.

An inhomogeneous and/or anisotropic space in general expands
on average differently than a space that is exactly homogeneous
and isotropic. This feature of general relativity is known as backreaction
\cite{Shirokov:1962, Buchert:1995, fitting, Buchert:1999a, Buchert:2001}; see
\cite{Ellis:2005, Buchert:2007, Rasanen:2010b, Rasanen:2011a, Buchert:2011}
for reviews.
The possibility that the change of the expansion rate due to
structure formation would explain the
observations of longer distances and faster expansion
is called the backreaction conjecture
\cite{Buchert:2000, Wetterich:2001, Schwarz:2002, Rasanen, Kolb:2004}.
Because of backreaction, the average expansion rate can
accelerate even in a dust universe in which the local expansion
rate decelerates everywhere
\cite{Rasanen:2006a, Rasanen:2006b, LTBacc, Kai:2006, Boehm:2013}.
Inhomogeneity also changes the relation between the expansion
rate and distance, even in a statistically homogeneous and
isotropic universe
\cite{Rasanen:2008b, Rasanen:2009b, Bull:2012, Boehm:2013},
so explaining the observations does not necessarily require
accelerating expansion.
In Newtonian cosmology \cite{Buchert:1995} and in relativistic
perturbation theory \cite{Rasanen:2011b} backreaction is small
(the issue has also been studied with a non-standard perturbative
formalism \cite{Green, Szybka:2013}).
Based on a gradient expansion, it has been argued that backreaction
is expected to be small even in the non-linear regime of structure
formation \cite{Rampf:2012}. However, as the density contrast goes
non-linear, gradients become large and the expansion breaks down.
In a semirealistic statistical model, the magnitude of the observed
change in the expansion rate and the timescale of ten billion years
emerge from the physics of structure formation \cite{Rasanen:2008a, peakrevs}.
There is no fully realistic calculation, and the amplitude
of backreaction in the real universe remains an open question.

\para{Expansion rate and light propagation.}

In studies of backreaction, the average expansion rate and other
spatial averages have often been considered
without relating them to observables
such as the redshift and the angular diameter distance.
On the other hand, in most work on light propagation in inhomogeneous
spacetimes it has been assumed, explicitly or implicitly,
that the average expansion rate is the same as in the
exactly homogeneous and isotropic Friedmann-Robertson-Walker (FRW)
model \cite{Rasanen:2008a}.

In \cite{Rasanen:2006b, Rasanen:2008a} it was suggested that in
statistically homogeneous and isotropic universes in which
the distribution evolves slowly, light propagation
over long distances could be described by a few average geometrical
quantities, namely the scale factor (or equivalently the average
expansion rate) and the average spatial curvature.
In \cite{Rasanen:2008b, Rasanen:2009b} it was argued
that redshift and angular diameter distance can, in a
dust spacetime, be approximately calculated
from just the average expansion rate and the value of the
matter density today (the average spatial curvature at all times
is not needed), and that null shear and light deflection are expected
to remain small.
Quantitative studies of light propagation were consistent
with this idea in the weak sense that in statistically homogeneous
and isotropic models in which the average expansion rate is close
to the FRW case, deviations in the redshift and distance
had been found to be small \cite{Rasanen:2008a}, and
this is the case also for later studies
\cite{SC, Bolejko, Bolejko:2010, Szybka:2010, Bolejko:2011, Bolejko:2012a, Clarkson:2011c}.
The small anisotropy of the cosmic microwave background
is also consistent with large deviation of the expansion
rate from the FRW case \cite{Rasanen:2009a}.

Recently, the average expansion rate and light propagation have
been considered in models that display either full or partial
statistical homogeneity and isotropy. In the absence of such
symmetry, there is no reason to expect the redshift and the
distance to be calculable from the average expansion rate.
Indeed, that is not the case in spherically symmetric models
\cite{Enqvist:2006, Bolejko:2008a, Kolb:2009, Bull:2012}.

Exact planar solutions with non-perturbative
inhomogeneities have been studied in \cite{Meures:2011, DiDio:2011}.
In \cite{Meures:2011}, the average expansion rate was not
calculated, but for the geometry considered, it reduces
to the FRW case when over- and underdensities are compensated
along the line of sight, and in this case the redshift and
the distance are also close to the FRW values.
In \cite{DiDio:2011}, variations in the expansion rate similarly cancel,
and results for light propagation are close to the FRW case.
In \cite{Bull:2012}, the relation between light propagation
and the average expansion rate was studied in several settings,
including a planar configuration with FRW and Kasner regions\footnote{A
somewhat similar setup was presented in \cite{Hellaby:2012}, with
spatially flat non-dust solutions glued together to produce a model
whose average expansion rate is different from FRW, but
light propagation was not considered.},
as well as a model with alternating expanding and collapsing
dust FRW regions. In contrast to previous studies,
the average expansion rate in the models is different
from the FRW case.
The average expansion rate was found to give a good
description of light propagation when the distribution along
the light ray is statistically homogeneous.
However, the model with alternating FRW regions (unlike the FRW-Kasner
model) is not a solution of the Einstein equation, as the
boundaries of different regions do not match together.
In that model, there is also some arbitrariness in the way the time
spent by the light ray in different regions was assigned.

The expansion rate and light propagation have also been
studied in models with discrete matter distribution
\cite{Clifton, lattice}\footnote{Some non-perturbative
calculations of the expansion rate, but not light propagation,
in the case of a distribution of discrete masses have also been
done in full numerical relativity \cite{BH}, and length scales in an
instantaneously static model were studied in \cite{Clifton:2012}.}.
The solutions of \cite{Clifton} are approximate
and the results depend on the method used to join
lattice cells together, with the more reliable method
leading to only a small correction to the FRW results.
In the perturbative calculation of \cite{lattice},
corrections to the FRW case were found to be small.
This is in agreement with the general result that
in the perturbative regime the average expansion rate
and the redshift are close to the FRW case. This is
likely also true for the distance if the universe is
statistically homogeneous and isotropic \cite{Rasanen:2011b}.
Calculations for the distance in first and second order in perturbation
theory agree with this conjecture \cite{Bonvin:2005, BenDayan:2012, BenDayan}.

These studies support the idea that the redshift and the angular diameter
distance are determined by the average expansion rate
when structures along the path of the light ray are a representative
sample of those in the volume over which the average is taken.
However, almost all exact solutions with statistical homogeneity
(at least along the light ray) have had average expansion rate close
to the FRW case, whereas models with average expansion rate
very different from FRW have not been exact solutions. One
exception is the Kasner-Eds model in \cite{Bull:2012}, though it
is not statistically isotropic.

We present the first statistically homogeneous and isotropic exact
solution in which inhomogeneity has a significant impact on the
expansion rate, and study the relation between the average
expansion rate and light propagation.
Our model is based on the Swiss Cheese construction
\cite{SC, Kai:2006, Bolejko, Szybka:2010, Marra:2011, Bolejko:2011}.
We start with a FRW dust model and replace some spherical
regions with the most general known exact dust solution,
the Szekeres model \cite{Szekeres}, \cite{Plebanski:2006} (page 387).
In \sec{sec:SSC} we prove that under certain general assumptions
the average expansion rate of the Szekeres Swiss Cheese model is
close to the background.
In \sec{sec:model} we present our cosmological model that violates
one of these assumptions, specialising to the spherically symmetric
subcase of the Szekeres model, the Lema\^{\i}tre-Tolman-Bondi (LTB) model
\cite{Plebanski:2006} (page 294), \cite{Lemaitre:1933, Tolman:1934, Bondi:1947}.
The model has surface layers, which have a large effect
on light propagation, and we consider modified versions
of the redshift and the angular diameter distance that are less
affected by the surface layers.
We study how well the redshift and the distance are described
by the average expansion rate, and consider effective equations of state.
In \sec{sec:disc} we discuss the results and their relation to previous
work. In \sec{sec:conc} we summarise our findings and mention
possible directions for extending the work.

\section{Szekeres Swiss Cheese} \label{sec:SSC}

\subsection{The Szekeres model}

\para{The metric, equations of motion and solutions.}

We consider inhomogeneous dust solutions (``holes'')
embedded in a FRW dust model (``background'' or ``cheese'').
We take the cosmological constant to be zero, so the
Einstein equation is
\bea \label{Einstein}
  G_{\a\b} = 8\pi\GN T_{\a\b} = 8\pi\GN \rhom u_\a u_\b \ ,
\eea

\noindent where $\GN$ is Newton's constant, $\rhom$ is the
dust energy density (we assume $\rhom\geq0$) and $u^\a=\delta^{\a}_{\ \, 0}$
is the four-velocity of observers comoving with the dust.
The solution can have surface layers, their contribution
is not included in \re{Einstein}.

The most general known exact solution of \re{Einstein}
is the Szekeres model \cite{Szekeres}, \cite{Plebanski:2006} (page 387).
The solution does not have any symmetries, \ie there are no Killing
vectors, but the form of the metric is nevertheless rather constrained.
In comoving synchronous coordinates the metric can be written as
\bea \label{metric}
  \rmd s^2 = - \rmd t^2 + X(t,r,p,q)^2 \rmd r^2 + \frac{R(t,r)^2}{\mE(r,p,q)^2} ( \rmd p^2 + \rmd q^2 ) \ ,
\eea

\noindent where $t$ is the proper time of observers comoving with
the dust fluid. Hypersurfaces of constant $t$ and $r$ are called shells.
The functions $X$ and $\mE$ are (we choose $X\geq0$)
\bea
  \label{X} X(t,r,p,q) &=& \frac{|R'-R \frac{\mE'}{\mE}|}{\sqrt{\epsilon+E(r)}} \\
  \label{E} \mE(r,p,q) &=& \frac{S(r)}{2} \left[ \left( \frac{p-P(r)}{S(r)} \right)^2 + \left( \frac{q-Q(r)}{S(r)} \right)^2 + \epsilon \right] \ ,
\eea

\noindent where $E(r)\geq-1, S(r), P(r)$ and $Q(r)$ are free functions,
prime denotes derivative with respect to $r$ and the parameter
$\epsilon$ takes on values $+1,0,-1$, respectively called
the quasispherical, quasiplanar and quasihyperbolic solutions.
When $\epsilon\geq0$, the coordinates $p$ and $q$ take values in the range
$]-\infty,\infty[$; see \cite{Szekeres} for the case $\epsilon=-1$.
The range of the $t$- and $r$-coordinates depends on the specific solution.
In the cases with $\epsilon\leq0$, the volume of the shells
is infinite, so because we are interested in holes with
finite volume, only the quasispherical solution with $\epsilon=+1$
is relevant. (We do not consider non-trivial topologies, which
could make the hypersurfaces compact.)
The function $R(t,r)\geq0$ satisfies the equations
\bea
  \label{eom} \Rdot(t,r)^2 &=& \frac{2 M(r)}{R(t,r)} + E(r) \\
  \label{rho} \rhom(t,r,p,q) &=& \frac{1}{4\pi\GN} \frac{ M'-3M\frac{\mE'}{\mE}}{ R^2 \left( R'-R\frac{\mE'}{\mE} \right) } \ ,
\eea

\noindent where dot denotes derivative with respect to $t$ and
$M(r)$ is a free function with the dimension of length.
The solutions of \re{eom} are as follows.

For $E(r)<0$,
\bea \label{solneg}
  R(t,r) = \frac{M}{|E|}(1-\cos\eta) \ , \quad
  \eta-\sin\eta = \frac{|E|^{3/2}}{M}(t-\tB(r)) \ .
\eea

For $E(r)=0$,
\bea \label{solzero}
  R(t,r) = \left(\frac{9}{2}M(t-\tB(r))^2\right)^{1/3} \ .
\eea

For $E(r)>0$,
\bea \label{solpos}
  R(t,r) = \frac{M}{E}(\cosh\eta-1) \ , \quad
  \sinh\eta-\eta = \frac{E^{3/2}}{M}(t-\tB(r)) \ .
\eea

\noindent Here $\tB(r)$ is a free function that
indicates the time when the big bang happens at radius $r$.
Time-reversed versions of \re{solneg}--\re{solpos}, where
$\tB(r)$ indicates the big crunch time instead, are also solutions.
In addition to the big bang or big crunch, Szekeres models can also
have shell crossing singularities
\cite{Szekeres, Hellaby:1985, Szekeres:1999}.
We consider only models that are free of shell crossings at least
up to the present day.

\para{Subcases.}

When $\mE'=0$, we have the isotropic subcase of the Szekeres model,
the LTB model
\cite{Plebanski:2006} (page 294), \cite{Lemaitre:1933, Tolman:1934, Bondi:1947}.
In this case, $R$ is the areal radius and
$R_\mathrm{p}\equiv\int_0^r\rmd r' X(t,r')\geq0$ is the proper
radius of a sphere centred on $r=0$. The function $X$ reduces to
\bea \label{LTBX}
  X(t,r) = \frac{|R'(t,r)|}{\sqrt{1+E(r)}} \equiv \frac{R'(t,r)}{\sqrt{1+E(r)}} s(r) \ ,
\eea

\noindent where $s(r)\equiv\sign(R')$ (any zeroes of $R'$ have to
be at constant $r$ for the density \re{rho} to be non-divergent).
For observers comoving with the dust fluid, the volume expansion rate
$\theta$, shear tensor $\sigma^{\a}_{\ \b}$, and spatial curvature $\sR$
are (for the definitions, see
\cite{Ehlers:1961, Ellis:1971, Ellis:1998c, Tsagas:2007, Magni:2012})
\bea \label{cov}
  \theta &=& 2 \frac{\Rdot}{R} + \frac{\Rdot'}{R'} \el
  \sigma^{\a}_{\ \b} &=& \diag\left( 0 , \frac{2}{3} , - \frac{1}{3} , - \frac{1}{3} \right) \left( \frac{\Rdot'}{R'} - \frac{\Rdot}{R} \right) \el
  \sR &=& - 2 \frac{(ER)'}{R^2 R'} + 4 \frac{\sqrt{1+E}}{R} \frac{1}{X} s' \ ,
\eea

\noindent and the shear scalar is
$\sigma\equiv\sqrt{\frac{1}{2}\sigma_{\a\b}\sigma^{\a\b}}$.
Vorticity is necessarily zero because of spherical
symmetry\footnote{Vorticity is also zero in the general Szekeres model.}
and four-acceleration is zero because the matter is dust.
If $R'$ changes sign, there is a surface layer signified by
the presence of the delta function $s'$, unless $E=-1$ at the
location where $R'=0$. 

When $E^3/M^2=$ constant and $\tB'=0$, we have the homogeneous subcase
of the LTB model (\ie the homogeneous and isotropic subcase of the
Szekeres model), the FRW dust model. Coordinates can be chosen such that
\bea \label{FRW}
  R(t,r) &=& a(t) r \ , \quad E(r) = - K r^2 \ , \quad M(r) = \frac{4\pi\GN}{3} \rhom(t) a(t)^3 r^3 \el
  \rhom(t) &=& \rhom(t_0) \frac{a(t_0)^3}{a(t)^3} \ , \quad
  \theta(t) = 3 H \ , \quad \sigma^{\a}_{\ \b} = 0 \ , \quad \sR(t) = 6 \frac{K}{a(t)^2} \ ,
\eea

\noindent where $K$ is a constant and $H\equiv\adot/a$. 
The subscript 0 refers to quantities evaluated at the present time.
The spatially flat case ($K=0$) is known as the
Einstein-de Sitter (EdS) model.

\para{Matching conditions.}

A solution can consist of different Szekeres regions matched on
hypersurfaces of constant $r$. The matching is smooth if the
Darmois junction conditions are satisfied \cite{junction}.
This means that the metric and the extrinsic curvature
are continuous. (See \cite{Matravers:2000} for an exhaustive
treatment of possible smooth matchings in the LTB case,
and  \cite{Mars:2013} for a review of the case with static holes.)
In particular, when $R'$ changes sign at one
or more shells $r=r_i$, the Darmois junction conditions
require $\mE'(r_i,p,q)=0$ and $E(r_i)=-1$.
If $E(r_i)\neq-1$, there is a surface layer at $r_i$,
and the extrinsic curvature has a finite jump whereas
$R$, $\mE$, $E$, $M$, and $\tB$ remain continuous and non-divergent
(see \cite{Hellaby:1985, Bonnor:1985, Humphreys:1998} for the LTB case).
The problem is not that $R'$ changes sign, but that $X$ has an absolute
value structure.
The sharp edge in the metric function $X$ implies that there is a delta
function contribution in the Einstein tensor and the Weyl tensor,
as in brane cosmology \cite{Shiromizu:1999}.
This surface layer can be viewed as a 2+1-dimensional submanifold
between regions of the 3+1-dimensional manifold that
have different signs of $R'$.
There is a corresponding delta function contribution on the
matter side of the Einstein equation, which can be interpreted as
a 2+1-dimensional energy-momentum tensor on the surface layer.
In the LTB case it corresponds to energy density
$^{(3)}\rho=\pm 32\pi\GN\sqrt{1+E}/R$ and pressure
$^{(3)}p=-\frac{1}{2} ^{(3)}\rho$. The sign is positive
if $R'$ switches from positive to negative as $r$ increases
past $r_i$ and negative in the opposite case\footnote{There is
a sign mistake in equation 19 of \cite{Humphreys:1998}.}.

\para{The average expansion rate.}

The volume of an inhomogeneous dust region does not necessarily
evolve like the volume of a homogeneous and isotropic dust model
\cite{Buchert:1995, Buchert:1999a}.
In other words, the average expansion rate can be
different from the FRW case, \ie there
can be significant backreaction. The average expansion rate can
even accelerate, as has been demonstrated in LTB models
\cite{LTBacc, Kai:2006}\footnote{Conditions for volume
acceleration in LTB models with $R'>0$ have been studied in
\cite{Sussman:2011}. However, the analysis seems to have errors.
In particular, the proof of Lemma 1 on page 10, quoted from
proposition 3 of their reference 54, is incorrect, because
parameters $\a, \b, \c, \d$ are not independent of $r$ and $r_\mathrm{tv}$.}.
However, embedding the Szekeres model into the FRW model constrains
the evolution. It is sometimes said that the average
expansion rate of a Swiss Cheese model is necessarily close to the
background FRW model, and approximate arguments to that effect
have been presented for certain types of LTB regions \cite{Marra:2011}.
We consider the issue for general quasispherical Szekeres regions.

The proper volume of a region on the hypersurface of constant
proper time $t$ with coordinate radius $r_V$ centred on the origin is
\cite{Bolejko:2008b}
\bea \label{V}
  V(t,r_V) = \int_V\rmd V = 4 \pi \int_0^{r_V} \rmd r \frac{ |R'| R^2}{\sqrt{1+E}} \ ,
\eea

\noindent where $\rmd V$ is the proper volume element. Note that
the function $\mE$ does not affect the volume.
The volume expansion rate can be written as $\theta=\rmd \dot V/\rmd V$.
The average of a scalar quantity $f$ on the hypersurface of constant
proper time $t$ in a domain with coordinate radius $r_V$ centred on
the origin is
\bea \label{av}
  \av{f} = \frac{\int_V\rmd V f}{\int_V\rmd V} \ ,
% for LTB = \frac{\int_0^{r_V} \rmd r \frac{ |R'| R^2 }{ \sqrt{1+E}} f }{ \int_0^{r_V} \rmd r \frac{ |R'| R^2}{\sqrt{1+E}} } \ .
\eea

\noindent  and the average expansion rate is $\av{\theta}=\dot V/V$.

\subsection{Swiss Cheese theorem} \label{sec:theorem}

\para{Theorem.}

Assume that the following conditions are satisfied for a Szekeres model
with vanishing cosmological constant.

\begin{enumerate}

\item There is a regular origin at $r=0$, so $R(t,0)=0$.

\item The function $R$ is monotonic in the coordinate $r$, $R'\geq0$.

\item There is a big bang singularity at $t=\tB(r)\geq0$, and there
are no other singularities at least until time $t=t_0$, with
$t_0-\tB(r)\sim t_0$.

\item The spacetime matches smoothly to a FRW dust universe (called
the background) at $r=\rb$.

\item At $t=t_0$, the function $R$ at the matching surface is small
compared to the background spacetime curvature radius.

\end{enumerate}

\noindent Then the average expansion rate is close to the background FRW value
\bea
  \av{\theta} \simeq 3 H \left[ 1 + \sO(\e) \right] \ ,
\eea

\noindent at all times $t\leq t_0$, where
$\e\sim\max\left\{ (\rb/t_0)^2, \left[\rb/(t_0-\tB)\right]^{\frac{2}{3}} \left( H_0^2 \rb^2 + K^2 \rb^2 \right)^{\frac{2}{3}} \right\}$.
Without loss of generality, we have chosen
$\tB(r)=0$ and $a(t_0)=1$ for the background.

\para{Proof.}

We first demonstrate that if $|E(r)|\ll1$ for all $r\leq\rb$,
the average expansion rate is close to the FRW background.
We then show that the above assumptions imply $|E|\ll1$.
If we have $|E|\ll1$, the volume of the hole is 
\bea
  V(t, \rb) &=& 4\pi \int_0^{\rb} \rmd r \frac{ |R'| R^2}{\sqrt{1+E}} \el
  &\simeq& 4\pi [ 1 + \sO(\Em) ] \int_0^{\rb} \rmd r R' R^2 \el
%  &=& \frac{4\pi}{3} R(t,\rb)^3 [ 1 + \sO(\Em) ] \el 
  &=& \frac{4\pi}{3} a(t)^3 \rb^3 [ 1 + \sO(\Em) ] \ ,
\eea

\noindent where $\Em$ is the maximum value of $|E|$ within $r\leq\rb$.
As the relative deviation of the volume from the volume of a FRW background
region (with the same coordinate radius) is small and independent of time,
the average expansion rate $\av{\theta}=\dot V/V$ is also close to
its background value.

Let us now prove that the condition $|E|\ll1$ holds.
There can be regions inside the hole with different signs
of $E$. We first consider regions with $E>0$. From \re{eom} we have 
\bea
  \Rdot &=& \pm \sqrt { \frac{2 M}{R} + E } \ .
\eea

\noindent The function $\mE'$ has at least one zero for all $r$,
so the requirement that the energy density is non-negative
together with assumption 2 implies that $M'\geq0$ via \re{rho}.
A regular origin implies that $M(0)=0$, so $M\geq0$.
Therefore $\Rdot$ cannot vanish for a shell with $E>0$, so it is
either positive or negative at all times. The latter case
implies that the region has always collapsed, in contradiction
with assumption 3. We thus have $\Rdot>0$ and
\bea \label{explimit}
  R(t,r) &=& R(0,r) + \int_0^t \rmd t \sqrt{\frac{2 M(r)}{R(t,r)} + E(r)} \geq t \sqrt{E(r)} \ ,
\eea

\noindent so we get
$E\leq R(t,r)^2/t^2\leq R(t,\rb)^2/t^2=a(t)^2 \rb^2/t^2$.
Putting $t=t_0$ and using assumption 5, we have $E\ll1$.
The physical reason is that positive $E$ increases the expansion
rate, and because the hole is small, a shell that expands too
fast will soon encounter the boundary of the hole,
leading to a shell crossing singularity.

Let us now consider regions with $E<0$. Equation \re{solneg} shows
that they collapse at $\eta=2\pi$, so
\bea \label{collimit}
  t-\tB(r) &<& \frac{2\pi M(r)}{|E(r)|^{\frac{3}{2}}} \leq \frac{2\pi M(\rb)}{|E(r)|^{\frac{3}{2}}}
%  &=& \frac{8\pi^2\GN}{3} \bar\rhom a^3 \rb^3 \frac{1}{|E(r)|^{\frac{3}{2}}} \el
  = \pi \left( H^2 + \frac{K}{a^2} \right) a^3 \rb^3 \frac{1}{|E(r)|^{\frac{3}{2}}} \ ,
\eea

\noindent where the second inequality follows from $M'\geq0$
and the last equality follows from \re{eom} and \re{FRW}.
We get the inequality
$|E|<\pi^{\frac{2}{3}}\left( \frac{a \rb}{t-\tB}\right)^{\frac{2}{3}} \left( H^2 a^2 \rb^2 + K^2 \rb^2 \right)^{\frac{2}{3}}$,
so putting $t=t_0$ and using assumption 5 gives $|E|\ll1$.
In this case the physical reason is that regions with more
negative $E$ expand slower and collapse sooner, so large $|E|$
corresponds to a short-lived region. This concludes the proof.

\para{Comments.}

In addition to excluding models with shell crossings, assumption 3
rules out geodesically extendible backgrounds and backgrounds that are
time-reverse of models that expand from a big bang singularity.
The condition $t_0-\tB(r)\sim t_0$ means that at $t_0$
the age of all regions of the universe is of the order of the
age of the background universe.
Assumption 4 could be replaced by the weaker condition
that the spacetime approaches the background solution only asymptotically,
with the deviation of the functions $R$ and $M$ from the background being
small at $r=\rb$.
Regarding assumption 5, the spacetime curvature radius of the background is
$|R^{0}_{\ 0}|^{-\frac{1}{2}} = \left(\frac{3}{2}\right)^{-\frac{1}{2}} ( H^2 + K/a^2 )^{-\frac{1}{2}}$.
For a spatially flat or negatively curved background, we have
$|K|/a^2\leq H\sim t^{-1}$, so the small radius condition
reduces to $\rb\ll H_0^{-1}\sim t_0$.
For a positively curved background, the condition is
$\sqrt{ \rb^2 H_0^2 + (\rb/r_K)^2 }\ll1$, where
$r_K\equiv K^{-1/2}$ is the maximum coordinate radius of the hypersphere.
So in addition to $\rb\ll H_0^{-1}$, the hole has to take up
a small portion of the total volume, $\rb\ll r_K$.
Together, these statements imply that also in this case $\rb\ll t_0$.
(Note that for fixed $\rb$, it is impossible to satisfy the condition
$a\rb\ll |R^{0}_{\ 0}|^{-\frac{1}{2}}$ at early times, regardless of the
value of $K$, because then $|R^{0}_{\ 0}|^{-\frac{1}{2}}\propto t$
and $a\rb\propto t^{2/3}$.)

Assumption 2, $R'\geq0$, can be replaced by assumption $2'$,
according to which there are no surface layers, \ie the Darmois
junction conditions are satisfied. This can be seen as follows.
Assume that $R'$ changes sign at least once, and
denote the largest value of $r$ where this happens by $\rmax$.
From the Darmois conditions it follows that $E(\rmax)=-1$.
Because we have $R'\geq0$ for $r\geq\rmax$, the above proof
shows that the shell at $\rmax$ is short-lived
if assumptions 4 and 5 hold, so assumption 3 is violated.

\section{Cosmological Tardis model} \label{sec:model}

\subsection{Tardis spacetime}

\para{Constructing the model.}

In order to to have significant backreaction in a Swiss Cheese model,
we have to violate at least one of the assumptions listed in
\sec{sec:theorem}. The most physical way to avoid singularities would
be to include a realistic treatment of what happens when shells cross
and shock fronts form or how collapse is stabilised by rotation,
pressure or velocity dispersion \cite{collapse}.
This would take us beyond the Szekeres model, which is based on
irrotational dust.
Keeping to the Szekeres model, and considering holes
that are at late times much smaller than the horizon and have
regular centres, we drop assumption 2 about monotonicity of $R$.

We restrict to the LTB subcase, $\mE'=0$, and
consider the simplest possibility, where $R'$ has one
maximum at $r=r_1$ and one minimum at $r=r_2>r_1$.\footnote{Some
common choices of coordinate system such as $R(t_0,r)\propto r$
or $M(r)\propto r^3$ exclude this possibility
and thus restrict the generality of the solution.
Even though the LTB metric and the equations of motion
are covariant under the transformation $r\rightarrow r'(r)$,
this does not imply that any one of the three functions
$E$, $M$ and $\tB$ could be set to any functional form
without loss of generality.}
We thus have $R'>0$ for $0\leq r<r_1$ and $r>r_2$
and $R'<0$ for $r_1< r<r_2$. (The proper radius
$R_{\mathrm{p}}$ and the proper volume \re{V} are monotonic in $r$.)
%While there are no indications of non-monotonic areal
%radius for realistic voids, the situation is not entirely
%unfamiliar. For an observer located in the centre,
%the areal radius is equal to the angular diameter distance $\DA$,
%and in many FRW models (including the EdS model) $\DA$ is a
%non-monotonic function of the redshift.
Equation \re{rho} shows that $M'$ has zeroes at the same values of $r$
as $R'$, so $M'$ is negative for $r_1<r<r_2$.
Nevertheless, the energy density integrated over the proper volume
is monotonic in $r$.
The function $M=4\pi\int \rmd r R' R^2 \rhom$ differs
from the volume integral of the energy density by the absence of
$1/\sqrt{1+E}$ in the integrand and, more importantly,
the substitution of $R'$ for $|R'|$.
The function $M(r)$ is the effective (or active)
gravitating mass, which generates the gravitational field
\cite{Plebanski:2006} (page 298), \cite{Bondi:1947}, and there
is no physical reason for it not to decrease with radius.
In fact, $\rhom\geq0$ does not even rule out $M<0$,
though we will have $M\geq0$ everywhere
($M<0$ would imply $\ddot{R}>0$).
For discussion of mass in general relativity
in a static setting, see \cite{Cederbaum:2012}.

\begin{figure}
%\hfill
\begin{minipage}[t]{7.7cm} 
\scalebox{1.0}{\includegraphics[angle=0, clip=true, trim=0cm 0cm 0cm 0cm, width=\textwidth]{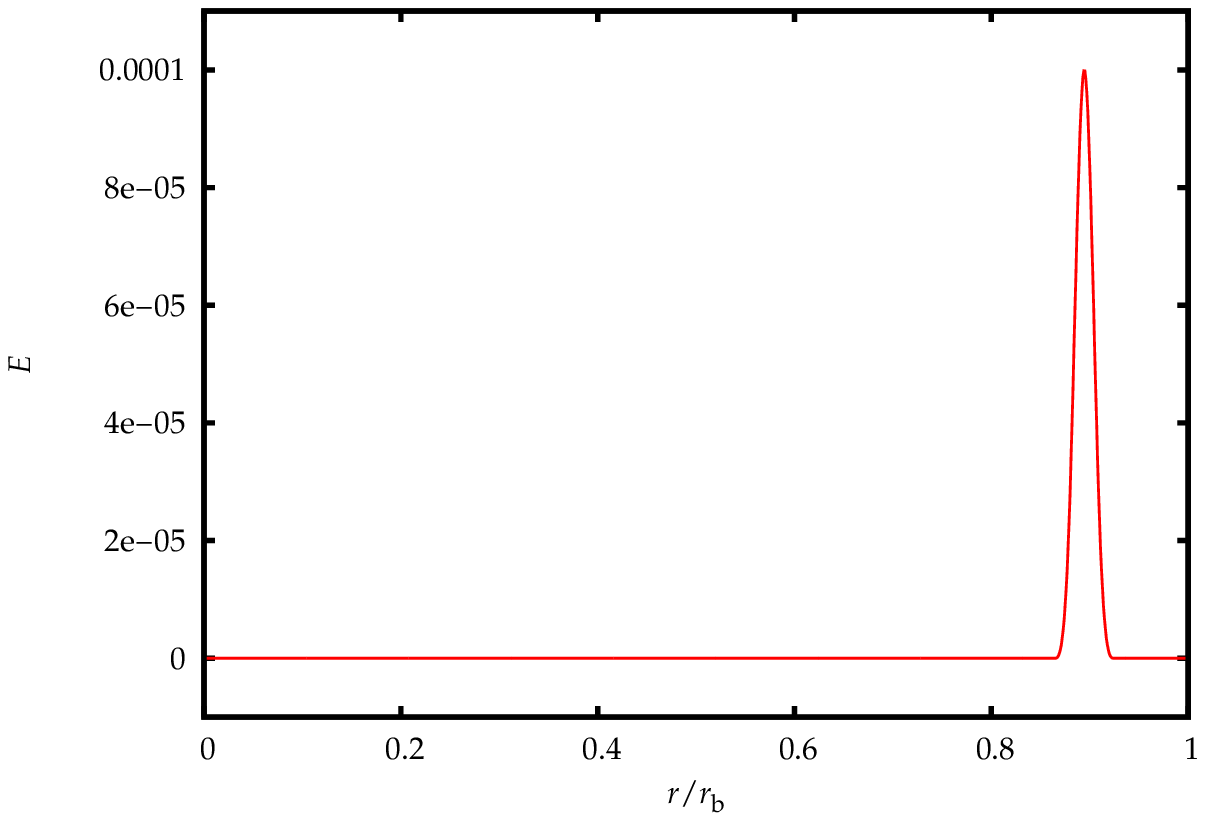}}
\begin{center} {\bf (a)} \end{center}
\end{minipage}
%\hfill
\begin{minipage}[t]{7.7cm}
\scalebox{1.0}{\includegraphics[angle=0, clip=true, trim=0cm 0cm 0cm 0cm, width=\textwidth]{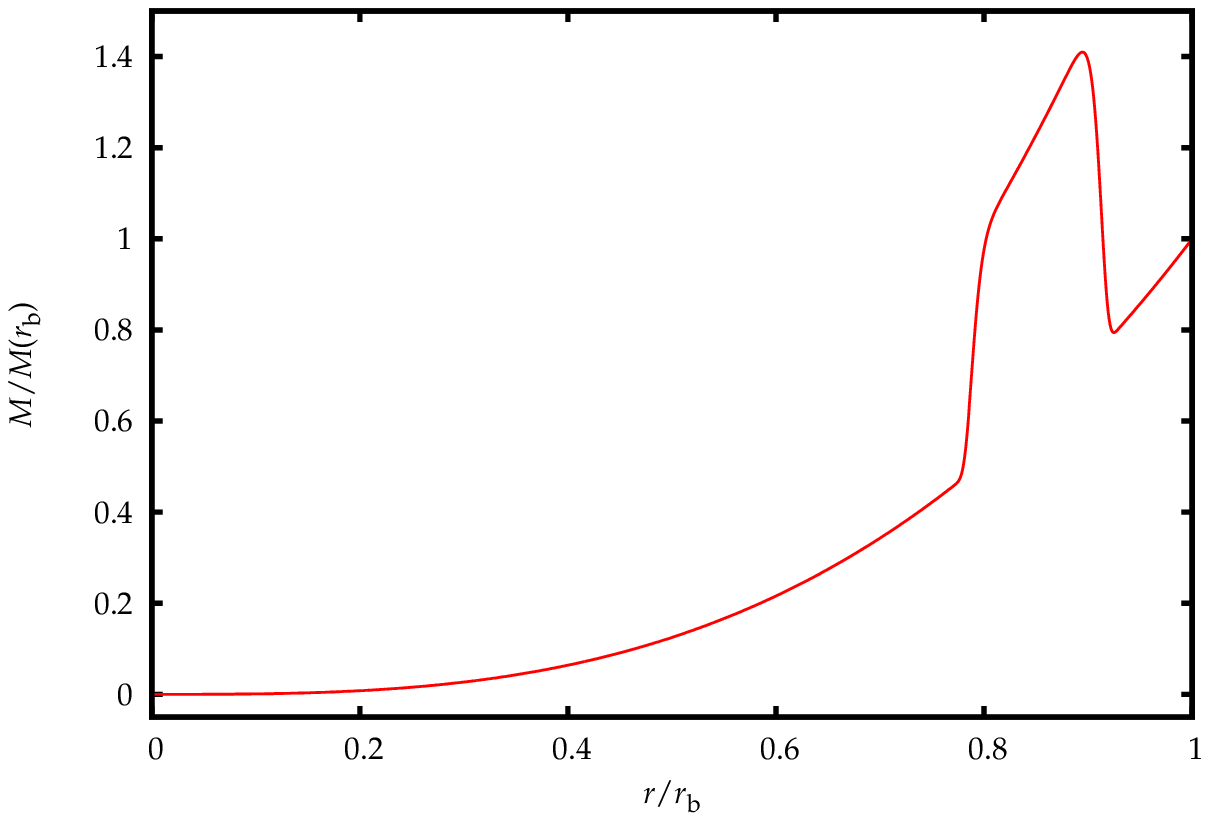}}
\begin{center} {\bf (b)} \end{center}
\end{minipage}
%\hfill
\caption{a) The function $E(r)$. b) The function $M(r)$ divided
by its value on the boundary, $M(\rb)$.}
\label{fig:e+m}
\end{figure}

\begin{figure}[ht!]
%\hfill
\begin{minipage}[t]{7.7cm} 
\scalebox{1.0}{\includegraphics[angle=0, clip=true, trim=0cm 0cm 0cm 0cm, width=\textwidth]{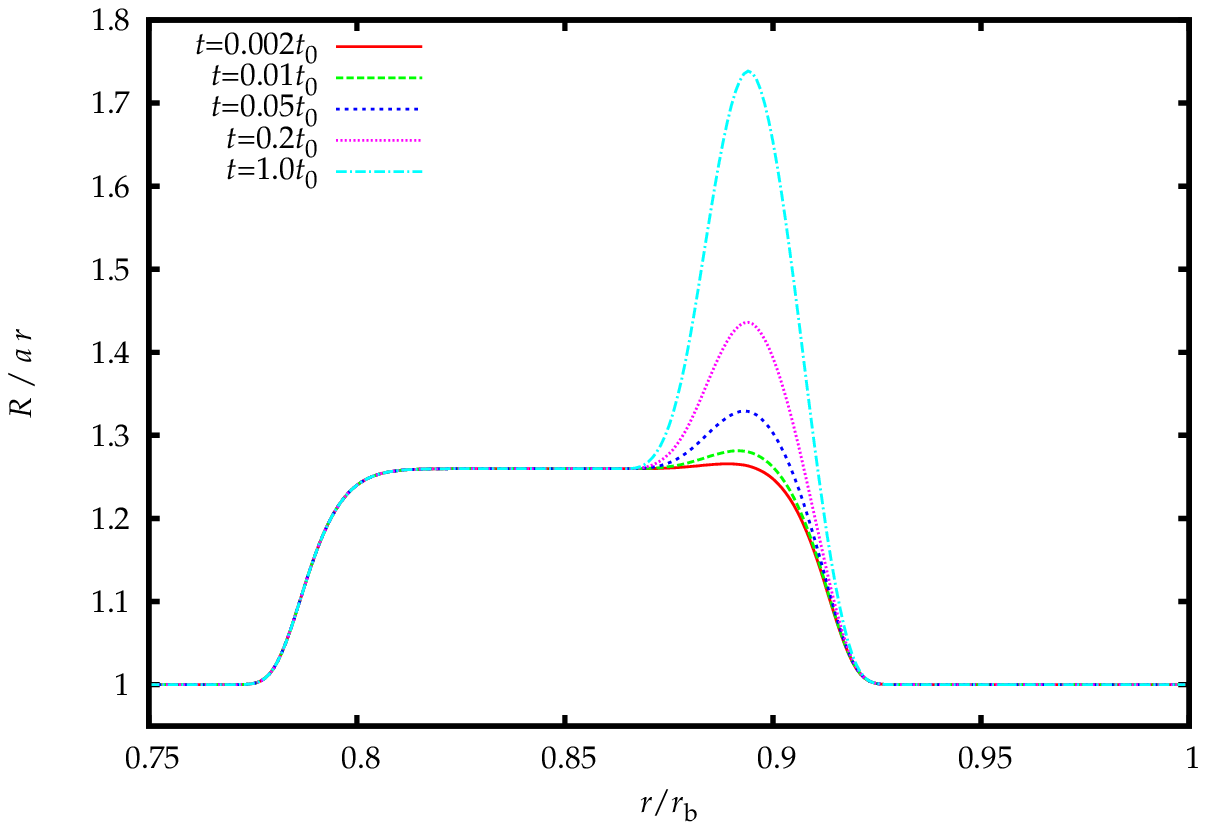}}
\begin{center} {\bf (a)} \end{center}
\hfill
\end{minipage}
%\hfill
\begin{minipage}[t]{7.7cm}
\scalebox{1.0}{\includegraphics[angle=0, clip=true, trim=0cm 0cm 0cm 0cm, width=\textwidth]{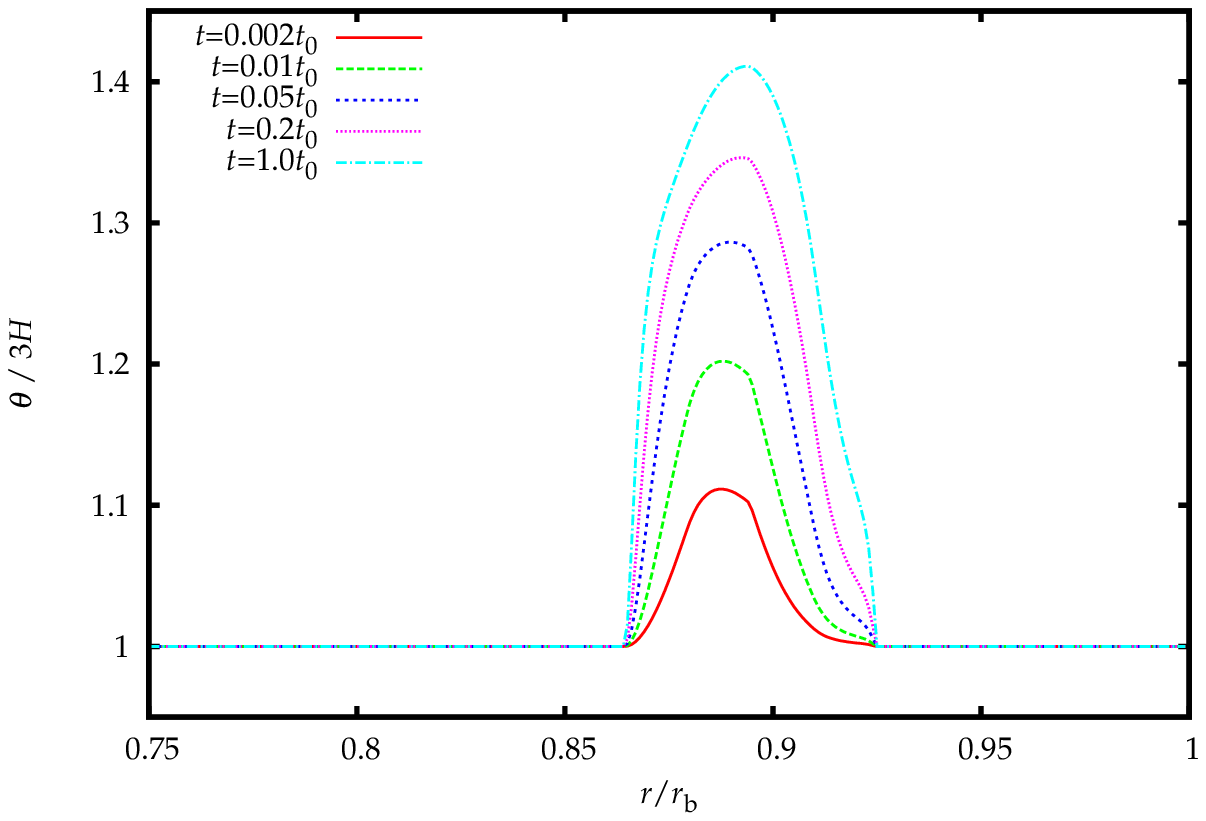}}
\begin{center} {\bf (b)} \end{center}
\hfill
\end{minipage}
%\hfill
\begin{minipage}[t]{7.7cm}
\scalebox{1.0}{\includegraphics[angle=0, clip=true, trim=0cm 0cm 0cm 0cm, width=\textwidth]{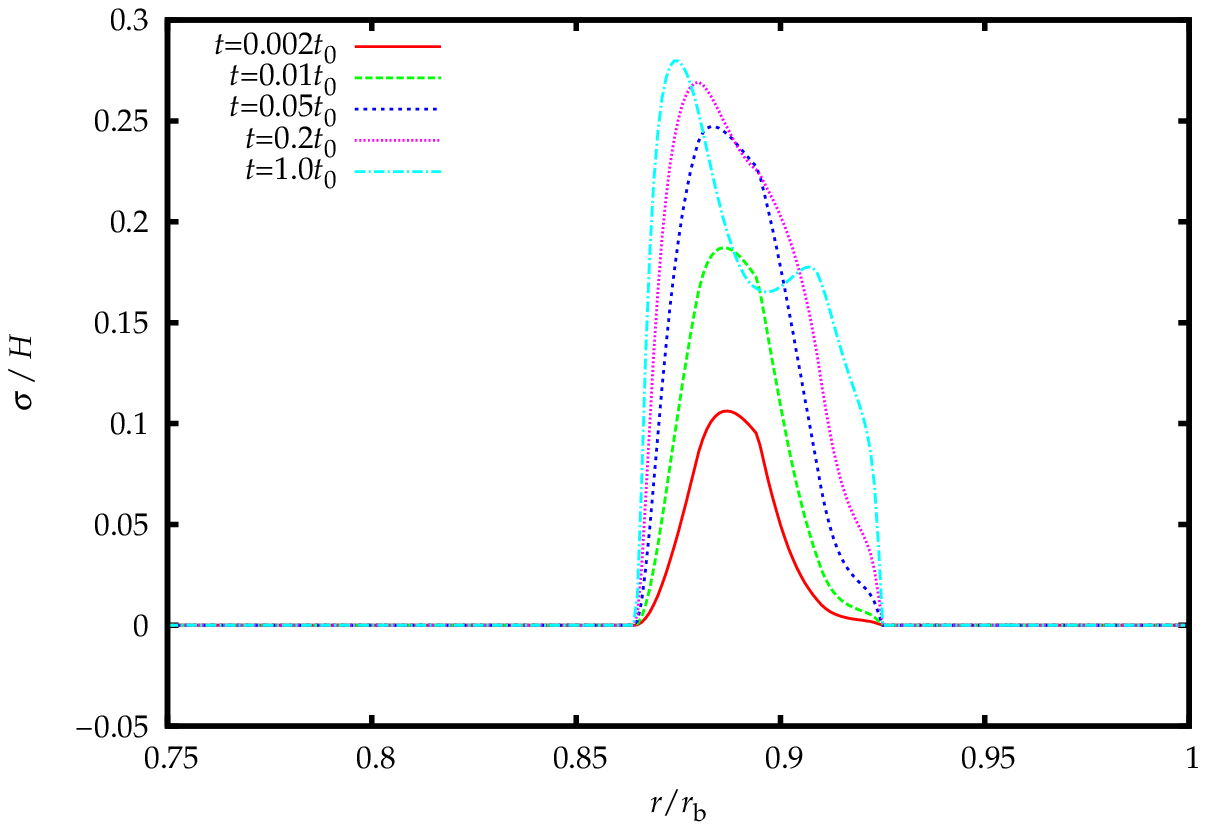}}
\begin{center} {\bf (c)} \end{center}
\end{minipage}
%\hfill
\begin{minipage}[t]{7.7cm}
\scalebox{1.0}{\includegraphics[angle=0, clip=true, trim=0cm 0cm 0cm 0cm, width=\textwidth]{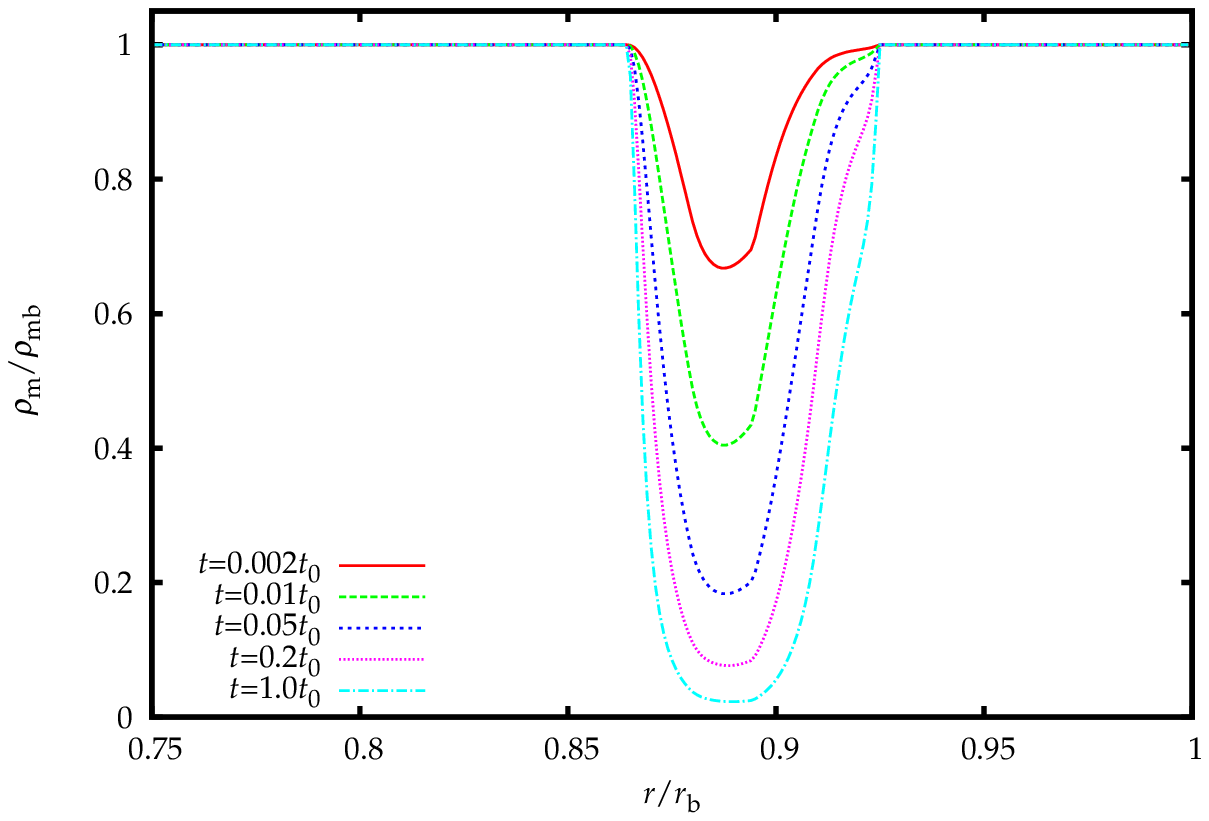}}
\begin{center} {\bf (d)} \end{center}
\end{minipage}
%\hfill
\caption{a) The areal radius relative to the background value,
$R/(a r)$.
b) The local expansion rate normalised to the background
expansion rate, $\theta/(3 H)$.
c) The shear scalar normalised to the background expansion rate, $\sigma/H$.
d) The energy density normalised to the background energy density,
$\rhom/\rho_\mathrm{mb}$.}
\label{fig:cov}
\end{figure}

With a non-monotonic $R'$, the condition $|E|\ll1$ is no longer
required to avoid early shell crossing or collapse.
Nevertheless, $E(r)\gtrsim1$ implies
$R(t_0,r)\gtrsim (t_0/\rb) \sqrt{E(r)} R(t_0,\rb)\gg R(t_0,\rb)$
via \re{explimit}, so the discrepancy between the volume of the
hole and the region it displaces is more extreme than
in the case $E\ll1$. In the case $E(r)<0$, $|E(r)|\sim1$, we have the constraint
$M(r)\gtrsim \e^{-3} M(r_b)$. To avoid late-time apparent horizons,
the condition for which is $R=2M$ \cite{Plebanski:2006} (page 311),
there must be a similarly drastic difference in the values of $R$
inside and outside. We consider only the case $E\geq0$ and keep $E\ll1$.
We take a spatially flat background, so the cheese consists of the EdS model.
We are free to choose the three functions $M(r)$, $E(r)$ and $\tB(r)$.
We take $\tB(r)=0$, so at early times the hole is close to the FRW
background \cite{Silk:1977}. Otherwise, we do not attempt to make the
holes realistic. For $M(r)$ we choose
\bea
  M(r) = \frac{1}{2} H_0^2 r^3 \big[ 1 + e^{ - \left( \frac{ r-r_{\mathrm{peak}} }{ A \rb } \right)^{10}} \big] \ ,
\eea

\noindent where $r_{\mathrm{peak}}=0.85\rb$ and $A=10^{-1.2}$.
The large power of $r$ in the exponent is chosen to damp the tail,
so that the junction conditions are satisfied with good numerical
accuracy at $r=\rb$.
The function $M(r)$ has extrema at $r_1=0.89\rb$ and $r_2=0.92\rb$.
In this case, the condition for no shell crossings reduces to 
the requirement that $E'=0$ at $r=r_1$ and $\sign(E')\sign(M')\geq0$
elsewhere.
For $E(r)$, we choose a cubic spline that is symmetric about
a maximum at $r_1=0.89\rb$ and has $E(r_2)=0$.
The functions $E$ and $M$ are shown in \fig{fig:e+m}.
As discussed in \sec{sec:SSC}, there are surface layers at $r=r_i$.

\begin{wrapfigure}[13]{r}[0pt]{0.5\textwidth}
%\begin{figure}
\scalebox{0.5}{\includegraphics[angle=0, clip=true, trim=0cm 0cm 0cm 0cm, width=1.0\textwidth]{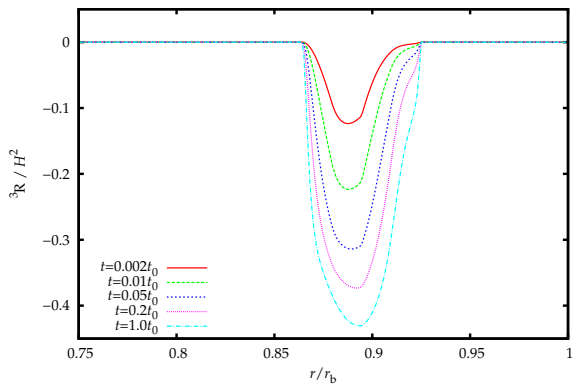}}
\caption{The spatial curvature normalised to the background
expansion rate, $\sR/H^2$.}
\label{fig:curv}
%\end{figure}
\end{wrapfigure}

In \fig{fig:cov} we show the areal radius, expansion rate, shear scalar
and energy density as a function of $r$ at different times.
The maximum value of the areal radius inside the hole reaches
about 1.7 times the background value, the minimum of the density drops to
less than 3\% of the background value, and the maximum of the
expansion rate grows to about 1.4 times the background value.
At early times, $\theta$, $\sigma$ and $\rhom$
asymptotically approach their FRW values.

The model is free of shell crossings at all times.
Because $R'$ changes sign, the areal radius inside the hole
can grow larger than the areal radius at the boundary, so it
is possible for shells to expand fast without colliding with
the boundary. The proper volume \re{V} of the hole is
\bea
  V(t, \rb) &\simeq& \frac{4\pi}{3} \left[ \Rb^3 + 2 R_1^3 - 2 R_2^3 \right] \ ,
\eea

\noindent where we have denoted $\Rb\equiv R(t,\rb), R_i\equiv R(t,r_i)$.
As $R_1>R_2$, the volume is larger  than in the FRW case, and the
average expansion (from \re{cov} and \re{av})
is correspondingly faster:
\bea
  \av{\theta} &\simeq& 3 H \frac{ 1 + 2 \frac{\Rdot_1 }{R_1} \left( \frac{R_1}{\Rb} \right)^3 - 2 \frac{\Rdot_2}{R_2} \left( \frac{R_2}{\Rb} \right)^3 }{ 1 + 2 \left( \frac{R_1}{\Rb} \right)^3 - 2 \left( \frac{R_2}{\Rb} \right)^3 } \ .
\eea

\noindent The average spatial curvature and the average
energy density are, from \re{rho}, \re{cov} and \re{av},
\bea
  \av{\sR} &\simeq& - 2 \frac{\int_0^{\rb} \rmd r \frac{ |R'|}{R'} (ER)' }{ \int_0^{\rb} \rmd r |R'| R^2 } = - 6 \frac{ E_1 R_1 }{ \Rb^3 + 2 R_1^3 - 2 R_2^3 } \el
  \av{\rhom} &\simeq& \frac{1}{4\pi\GN} \frac{ \int_0^{\rb} \rmd r |M'| }{ \int_0^{\rb} \rmd r |R'| R^2 } = \frac{3}{4\pi\GN} \frac{ M_{\mathrm{b}} + 2 M_1 - 2 M_2 }{ \Rb^3 + 2 R_1^3 - 2 R_2^3 } \ ,
\eea

\noindent where $E_i\equiv E(r_i), M_i\equiv M(r_i), M_{\mathrm{b}}\equiv M(\rb)$
and we have taken into account that absence of singularities
requires $E(0)=0, M(0)=0$.
Note that $|E|\ll1$ does not imply that the spatial curvature would
be small either locally or on average. The spatial curvature
is shown in \fig{fig:curv}. For our choices of $E$ and $M$, the
spatial curvature does not approach its FRW value even at early times,
unlike the expansion rate, shear and energy density.

\para{Inner and outer size.}

\begin{figure}
%\hfill
\begin{minipage}[t]{7.7cm} 
\scalebox{1.0}{\includegraphics[angle=0, clip=true, trim=0cm 0cm 0cm 0cm, width=\textwidth]{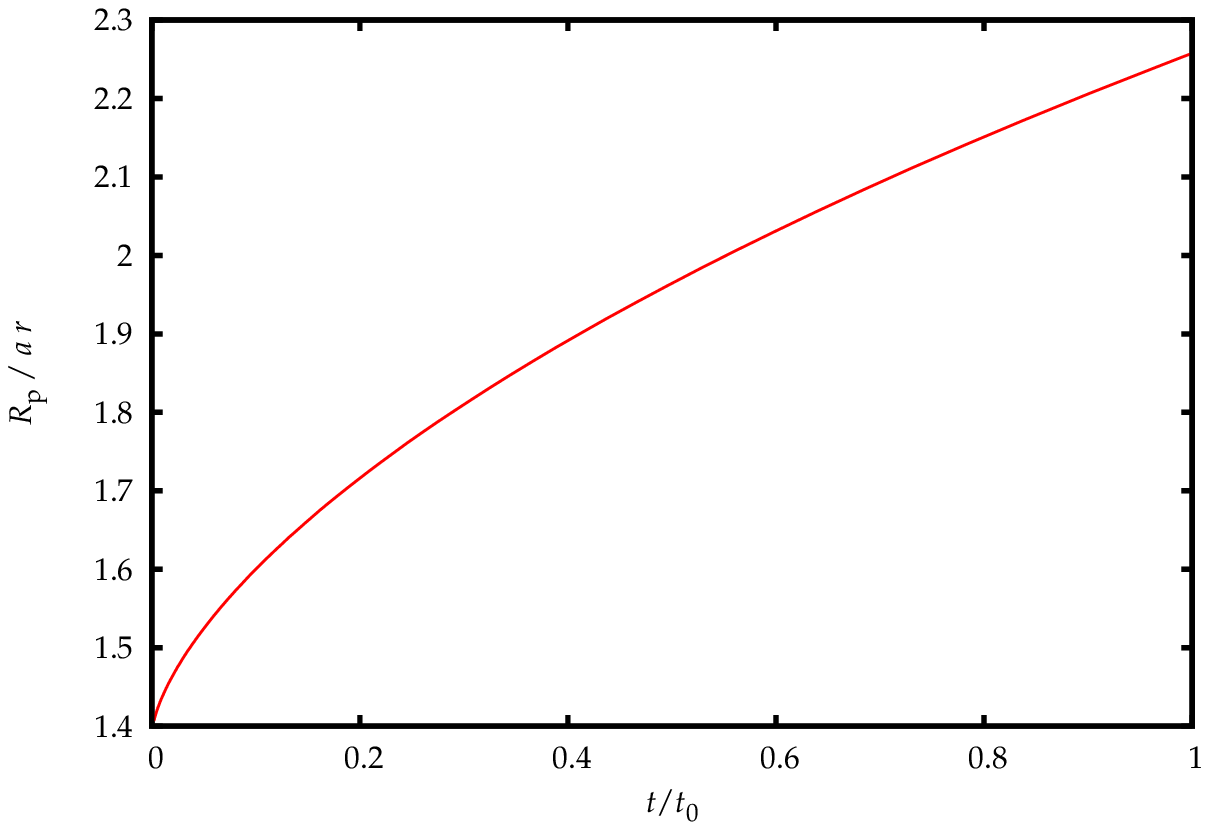}}
\begin{center} {\bf (a)} \end{center}
\end{minipage}
%\hfill
\begin{minipage}[t]{7.7cm}
\scalebox{1.0}{\includegraphics[angle=0, clip=true, trim=0cm 0cm 0cm 0cm, width=\textwidth]{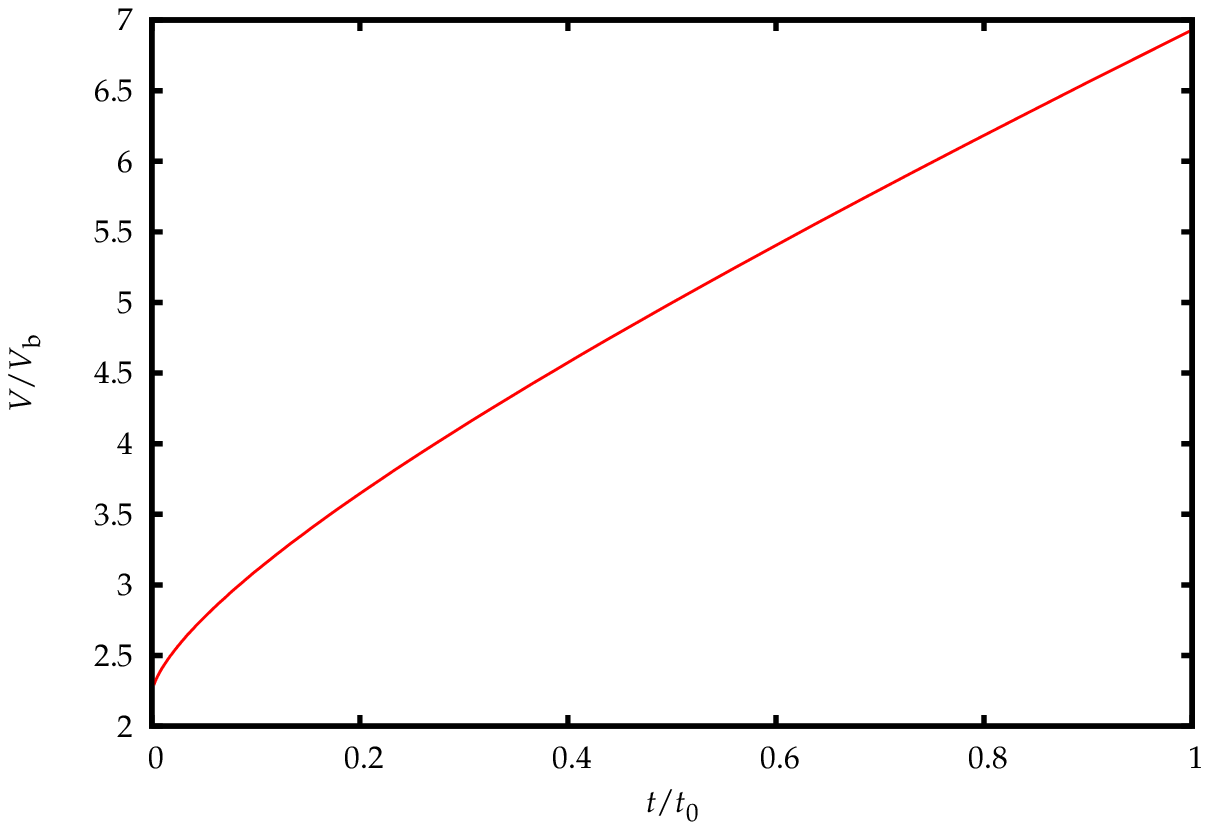}}
\begin{center} {\bf (b)} \end{center}
\end{minipage}
%\hfill
\caption{a) Proper radius relative to a background region with the same areal radius, $R_{\mathrm{p}}/(ar)$.
b) Proper volume relative to the volume of a background region with the same areal radius, $V/V_b$.}
\label{fig:V}
\end{figure}

We have removed a portion of spacetime and fitted in its
place another region that fits smoothly into the hole
on the boundary, but has larger spatial volume than the removed part.
We call a solution that features one or more such domains
whose inner dimensions seem at odds with the outer a
{\em Tardis spacetime}, and refer to the embedded domains as
{\em Tardis regions}\footnote{http://www.oed.com/viewdictionaryentry/Entry/247369}.
A Tardis region can be much larger than expected from its surface
area and the linear size it occupies in the background spacetime
based on Euclidean intuition.
(Note that the surface layer is located inside the
Tardis region, and the interface between the Tardis region
and the outside world appears normal.)
The proper radius and proper volume of the hole relative to
the background are shown in \fig{fig:V}.
The proper radius is today about 2 times as large as that
of the removed region, and the proper volume is about 7 times as large.
The relation $\av{\theta}=\dot{V}/V$ implies that
if the average expansion rate is different from the background
(and monotonic), the volume element will also be different, but the
reverse does not hold true. At early times, the average expansion rate is
close to FRW, but the proper radius and proper volume do not approach
their FRW values, as $R$ is discontinuous at all times.
It would be possible to tune the function $M(r)$ so that the
volume of the hole is close to the background at early times, but
this is not a necessary consequence of time evolution, unlike for
the expansion rate, shear and energy density.

As the volume element is defined locally,
its values inside a given volume are not determined by the
values on the boundary. We have defined the Tardis region in terms
of embedding into a background spacetime, so the volume of the
embedded region can be compared with the corresponding
background region. A realistic model would
not consist of isolated regions embedded in a homogeneous and
isotropic background that provides a global point of
comparison\footnote{Though as the real universe has been close to FRW
in the past, the volume of a region can be compared to the volume that
it would have if it had continued FRW evolution.}.
Nevertheless, the Tardis effect is central to general
relativity and can be formulated without reference to embedding.
Consider a spherical region (\ie a simply connected volume such
that all points on the boundary are at the same proper distance
from one point) in any three-dimensional curved space
and define the areal radius by $R\equiv\sqrt{S/(4\pi)}$, where $S$
is the area of the boundary. In general, the proper
radius of the sphere is not $R$ and its volume is not
$4\pi R^3/3$. A well-known example is given by spatially curved
FRW models. In the real universe, underdense regions with
negative curvature are in this sense larger than expected,
and overdense regions correspondingly smaller (neglecting
factors such as shear and rotation).

\begin{wrapfigure}[15]{r}[0pt]{0.5\textwidth}
%\begin{figure}
\scalebox{0.5}{\includegraphics[angle=0, clip=true, trim=0cm 0cm 0cm 0cm, width=1.0\textwidth]{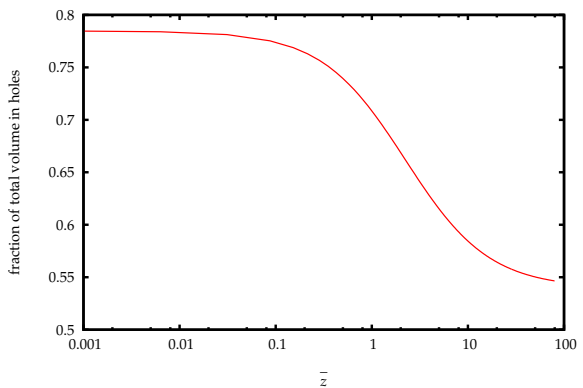}}
\caption{The fraction of volume taken up by the holes,
as a function of the mean redshift \re{bz}.}
\label{fig:volfrac}
%\end{figure}
\end{wrapfigure}

\para{Cosmological model.}

We construct a cosmological model with a distribution of time
evolving underdense Tardis regions. In the real universe voids,
underdense regions which expand faster than average,
are a central feature of the observed matter distribution.
The void distribution is dominated by voids of a certain
size, whose evolution is governed by the
power spectrum of perturbations.
Typical void radius today is of the order 20 Mpc $\sim 10^{-2} H_0^{-1}$
\cite{Sutter:2012}.
We consider holes that are all identical to each other,
with comoving areal radius $\rb=0.01 H_0^{-1}$,
so that their outer size is similar to that of real structures,
though their density profile (shown in \fig{fig:cov}) is not realistic.
Real cosmological structures have a complex
multiscale arrangement of clusters, filaments and walls in addition
to voids \cite{multiscale}, and modelling this accurately would
require dealing with the statistics of hierarchical structure
formation.

We distribute voids randomly in the background space with a
uniform distribution, and remove overlapping regions. The
voids take up about 34\% of the volume as measured by their
``outer size", \ie the size of the FRW regions they displace.
As the volume of the voids grows faster than the background, their
fraction of the real volume increases in time, as shown in \fig{fig:volfrac}.

\subsection{Light propagation} \label{sec:light}

\para{Redshift, angular diameter distance and null shear.}

We consider light propagation in the geometrical optics approximation,
so light travels on null geodesics
\cite{Schneider:1992} (page 93), \cite{Sasaki:1993}.
Photon momentum, which is tangent to the null geodesic,
is denoted by $k^\a=\frac{\rmd x^\a}{\rmd\l}$, where $\l$ is
the affine parameter.
Redshift measured by an observer comoving with the dust
is given by the ratio of the energy $E_{\mathrm{s}}$
emitted at the source and the observed energy $E_{\mathrm{o}}$,
\bea \label{z}
  1 + z &=& \frac{E_{\mathrm{s}}}{E_{\mathrm{o}}} = \frac{k^0}{k^0_{\mathrm{o}}} \ ,
\eea 

\noindent where we have used $E=-u_\a k^\a=k^0$. We decompose $k^\a$
into an amplitude and the direction, and split the direction into
components orthogonal and parallel to the dust geodesics,
\bea \label{kdec}
  k^\a = E ( u^\a + e^\a ) \ ,
\eea

\noindent where $u_\a e^\a=0$, $e_\a e^\a =1$.
In terms of the covariant quantities \re{cov}, the redshift is
(see e.g. \cite{Rasanen:2008b})
\bea \label{zcov}
  1 + z &=& \exp\left( \int_{\l_\mathrm{s}}^{\l_\mathrm{o}} \rmd\l E \left[ \frac{1}{3}\theta + \sigma_{\a\b} e^\a e^\b \right] \right) \ .
\eea

Because of spherical symmetry we can restrict our consideration
to the hypersurface $\theta=\pi/2$ without loss of generality,
so $k^\theta=0$. The other components $k^\a$ can be solved from the null
condition $k_\a k^\a=0$ and the null geodesic equation $k^\b\nabla_\b k^\a=0$.
From the $\phi$-component of the latter we get $k^\phi=\cp/R^2$, where
$\cp$ is a constant. Defining the Euclidean impact parameter as
$b\equiv \rmin^E/\rb$, where $\rmin^E$ is the smallest $r$-coordinate
the light ray would have were the space Euclidean, we have $\cp=b\rb$.
The other components of the null geodesic equation give
\bea
% \frac{\rmd k^0}{\rmd\lambda} + \frac{\Xdot}{X} (k^0)^2 + \left( \frac{\Rdot}{R} - \frac{\Xdot}{X} \right) \frac{\cp^2}{R^2} &=& 0 \el
 \label{geodesic0} \frac{\rmd k^0}{\rmd\lambda} + \frac{\Rdot'}{R'} (k^0)^2 + \left( \frac{\Rdot}{R} - \frac{\Rdot'}{R'} \right) \frac{\cp^2}{R^2} &=& 0\\
 \label{geodesicr} \frac{\rmd (X k^r)}{\rmd\lambda} + \frac{\Rdot'}{R'} k^0 X k^r - \frac{1}{X} \frac{R'}{R} \frac{\cp^2}{R^2} &=& 0\ .
\eea

\noindent The component $k^r$ generally diverges when $X=0$, but $X k^r$
remains finite. However, the derivative $\frac{\rmd (X k^r)}{\rmd\lambda}$
jumps because of the absolute value structure of $R'/X$ in the last
term in \re{geodesicr}. Correspondingly, the derivative of $k^\phi$
jumps, as the components are related by the null condition,
\bea \label{null}
  X k^r = \pm \sqrt{ (k^0)^2 - \frac{\cp^2}{R^2} } \ .
\eea

The area and shape of a bundle of null geodesics
are solved from the Sachs equations:
\bea \label{Sachs}
  \patl\thetat + \frac{1}{2} \thetat^2 + 2 \sigmat^2 &=& - G_{\a\b} k^\a k^\b \el
%  &=& - 8\pi\GN T_{\a\b} k^\a k^\b - 2 \sigmat^2 - \frac{1}{2} \thetat^2 \el
  \htt_{\a}^{\ \ \c} \htt_{\b}^{\ \ \d} \patl{\sigmat_{\c\d}} + \thetat \sigmat_{\a\b} &=& - k^\mu k^\nu \htt_{\a}^{\ \ \c} \htt_{\b}^{\ \ \d} C_{\mu\gamma\nu\d} \ ,
\eea

\noindent where $\thetat\equiv\nabla_\a k^\a$ is the area
expansion rate of the light bundle,
$\sigmat_{\a\b}\equiv\htt_{\a}^{\ \ \c} \htt_{\b}^{\ \ \d} \nabla_\d k_\c - \frac{1}{2} \htt_{\a\b} \thetat$
is the null shear tensor which describes image deformation,
$\sigmat\equiv\sqrt{\frac{1}{2}\sigmat_{\a\b}\sigmat^{\a\b}}$
is the null shear scalar, $\htt_{\a\b}$ projects on a
two-dimensional surface orthogonal to the light ray
and $C_{\a\b\c\d}$ is the Weyl tensor.
The quantities $\thetat$ and $\sigmat$ are independent of the
choice of $\htt_{\a\b}$.

The tensors $G_{\a\b}$ and $C_{\a\b\c\d}$ have
delta functions in the shells where $X=0$,
coming from terms of the form $X'/X$.
Exactly the same delta functions appear on the
left-hand side in $\patl\thetat$ and $\patl{\sigmat_{\a\b}}$,
so the delta function parts of the equations are satisfied
identically, and we drop them in what follows.
Because of spherical symmetry, $\sigmat_{\a\b}$ has only one independent
component, which we can express in terms of $\sigmat$.
For the LTB metric given by \re{metric} and \re{X}, the smooth parts
of the Sachs equations \re{Sachs} reduce to\footnote{In
\cite{Szybka:2010} there is a factor of 2 missing on the right-hand
side of the equation corresponding to \re{sigmaeq}. This has
little effect on the results, as the null shear is negligible.}
\bea
  \label{thetaeq} \patl\thetat + \frac{1}{2} \thetat^2 + 2 \sigmat^2 &=& - 8\pi\GN \rhom (k^0)^2 \\
  \label{sigmaeq} \patl\sigmat + \thetat \sigmat &=& \frac{\cp}{R^2} \left( 4\pi\GN \rhom - 3 \frac{M}{R^3} \right)  \ ,
\eea

\noindent where we have used the Einstein equation \re{Einstein}.
The area expansion rate is related to the angular diameter distance by
$D_A\propto \exp \left( {\frac{1}{2} \int \rmd\l \thetat} \right)$.
For an observer located in the centre, we would simply have
$\sigmat=0$ and $D_A=R$. In terms of $\DA$, \re{thetaeq} reads
\bea \label{DAl}
  \frac{\rmd^2 D_A}{\rmd\l^2} &=& - [ 4 \pi\GN \rhom (k^0)^2 + \sigmat^2 ] D_A \ .
\eea

The sources on the right-hand side of the Sachs equations
\re{thetaeq} and \re{sigmaeq} are continuous.
Therefore the equations have solutions that are continuous;
however, the solutions that correspond to the spacetime with surface
layers are those in which $\thetat$ and $\sigmat$ and their
first derivatives have finite discontinuities at the locations where $X=0$.
The magnitude of the jumps is given by the inverse size of the
hole, so for one hole the relative change is of order unity.
When travelling cosmological distances, the relevant scale
for $\thetat$ is $H$, so the jumps are much larger than
the continuous part.
In other words, the surface layers completely distort light
propagation compared to the smooth case. It is also not clear
whether the geometrical optics approximation holds, as the
curvature changes on an infinitely small scale when crossing
surface layers; the issue would have to be settled by studying
the thin limit of a thick shell.

We consider two modified versions of the light propagation
calculation. In the first case, we still calculate the photon
momentum from the null geodesic equation but modify the distance
calculation by picking the continuous solutions
of the Sachs equations \re{thetaeq} and \re{sigmaeq}
instead of the discontinuous solutions.
This corresponds to angular diameter distance in a spacetime
without the surface layers.
The absolute value structure of $X$ still has an effect
on the calculation by causing a jump in $\rmd k^\phi/\rmd\l$
according to the null geodesic equation \re{geodesicr}. This makes the
light ray turn sharply when it crosses a surface layer.
In order to assess the importance of this effect on the redshift,
we consider a second modification of light propagation,
in which we take the light rays to follow straight null lines
as defined by the background metric and calculate the redshift
from \re{zcov}.
This removes light deflection due to the unphysical surface layers.
(In solutions without surface layers, there would of course still
be some light deflection, but it would be small.)
However, these straight null rays are no longer geodesic, so
the Sachs equations do not hold and it is not meaningful to
calculate the angular diameter distance along them, so in the
second case we consider only the redshift.

\para{Light propagation calculation.}

We consider an octant of a ball with radius $60\rb$ centered on the
observer, and fill it with voids drawn from a uniform distribution,
removing overlapping spheres.
Beyond this region we generate the distribution of voids dynamically
along the null geodesic in order to cut down on computation time.
We integrate from the observer to the source, and stop the
calculation when we reach time $t=2\times 10^{-3} t_0$.
A typical light ray goes through approximately $35\pm5$ holes
for the case when the light rays are geodesic, and $40\pm5$ when
they go straight as defined by the background.
We propagate the null shear consistently using the Sachs
equations, using the same procedure as in \cite{Szybka:2010}, which
presented the first correct treatment of the shear in Swiss Cheese models
(in the prescription in which the light rays are propagated
from the observer to the source, as is common in the literature).
As in \cite{Szybka:2010}, the null shear is small, its contribution
relative to the density in \re{DAl} is at most $10^{-3}$.

Equation \re{DAl} is linear, so the normalisation of $D_A$ is
arbitrary. In the FRW model, the normalisation is fixed and
for small redshifts we have $D_A\simeq H_0^{-1}z$,
where $H_0$ is the current value of the Hubble parameter.
When the expansion rate is inhomogeneous, as in our model and
in the real universe, the correct normalisation is less clear.
The naive idea of using the volume expansion rate at the observer's
location is inappropriate. For an observer located in a stable object
such as a galaxy, the local expansion rate is zero, and
for an observer located in a collapsing region it is negative.
The choice of normalisation should be related to
how real observations are analysed.
In practice, the value of $H_0$ that is used is a
weighted average over some redshift range.
For example, the determination of $H_0$ in \cite{Riess:2011}
uses data in the ranges $0.023<z<0.1$ and $0.01<z<0.1$.
On the theoretical side, the issue has been discussed in
\cite{BenDayan:2012, Marra:2012}.
We normalise to the spatially averaged value of the volume
expansion rate, as also done in \cite{Bull:2012}.
Let us discuss spatial averages in more detail 
before showing the results of the light propagation
calculation and commenting on how well the redshift
and the distance are described by the average expansion rate.

\subsection{Average quantities} \label{sec:av}

\para{The average expansion rate.}

As discussed in \sec{sec:intro}, it has been suggested that
light propagation in a statistically homogeneous
and isotropic space in which the distribution evolves slowly
can be described in terms of the average expansion rate
\cite{Rasanen:2006b, Rasanen:2008a, Rasanen:2008b, Rasanen:2009b, Bull:2012}.
We can now test this idea with our Swiss Cheese model.

The evolution and constraint equations for a general geometry
can be written in terms of the covariant quantities \re{cov}.
For a general irrotational dust model, the scalar parts are
\cite{Ehlers:1961, Ellis:1971, Ellis:1998c, Tsagas:2007, Magni:2012}
\bea
  \label{Rayloc} \dot{\theta} + \frac{1}{3} \theta^2 &=& - 4\pi\GN \rhom - 2 \sigma^2 \\
  \label{Hamloc} \frac{1}{3} \theta^2 &=& 8\pi\GN \rhom - \frac{1}{2} \sR + \sigma^2 \\
  \label{consloc} 0 &=& \dot\rho_{\mathrm{m}} + \theta \rhom \ .
\eea

In the present case, there are surface layers in addition
to dust. Equation \re{cov} shows that $\theta$ and $\sigma$
are smooth everywhere.
In contrast, $\sR$ has a delta function contribution at the
locations where $R'=0$, as discussed in \sec{sec:SSC}.
We have not written down the surface layer contribution to the spatial
curvature in \re{Hamloc}, as it is exactly cancelled by the energy
density of the surface layer.

The scale factor $\ba(t)$ that describes the evolution
of the proper volume (not to be confused with the background
scale factor $a(t)$) is defined as
\bea
  \ba(t) \equiv \left( \frac{V(t)}{V(t_0)} \right)^{\frac{1}{3}} \ ,
\eea

\noindent and the average Hubble parameter is defined as
$\bH\equiv \badot/\ba$, or equivalently as $3\bH=\av{\theta}$.
Averaging \re{Rayloc}-\re{consloc} over the hypersurface
of constant $t$, we obtain the Buchert equations \cite{Buchert:1999a}
that describe the evolution of $\ba$:
\bea
  \label{Ray} 3 \frac{\baddot}{\ba} &=& - 4 \pi\GN \av{\rhom} + \sQ \\
  \label{Ham} 3 \frac{\badot^2}{\ba^2} &=& 8 \pi \GN \av{\rhom} - \frac{1}{2}\av{\sR} - \frac{1}{2}\sQ \\
  \label{cons} 0 &=& \pat_t \av{\rhom} + 3 \frac{\badot}{\ba} \av{\rhom} \ ,
\eea

\noindent where the backreaction variable is defined as
$\sQ\equiv\frac{2}{3}\left( \av{\theta^2} - \av{\theta}^2 \right) - 2 \av{\sigma^2}$.
The integrability condition between \re{Ray} and \re{Ham} is
\bea \label{int}
  && \pat_t \av{\sR} + 2 \bH \av{\sR} = - \dot\sQ - 6 \bH \sQ \ .
\eea

\noindent Dividing \re{Ray} and \re{Ham} by $3 \bH^2$,
we get \cite{Buchert:1999a}
\bea \label{omegas}
  \label{q} q_H &\equiv& - \frac{1}{\bH^2} \frac{\baddot}{\ba} = \frac{1}{2} \Om + 2 \OQ \\
  \label{Omegas} 1 &=& \Om + \OR + \OQ \ ,
\eea

\noindent where $\Om\equiv 8\pi G_N \av{\rhom}/(3 \bH^2)$,
$\OR\equiv-\av{\sR}/(6 \bH^2)$ and $\OQ\equiv-\sQ/(6 \bH^2)$
are the density parameters of matter, spatial curvature
and the backreaction variable, respectively.

\begin{figure}
%\hfill
\begin{minipage}[t]{7.7cm} 
\scalebox{1.0}{\includegraphics[angle=0, clip=true, trim=0cm 0cm 0cm 0cm, width=\textwidth]{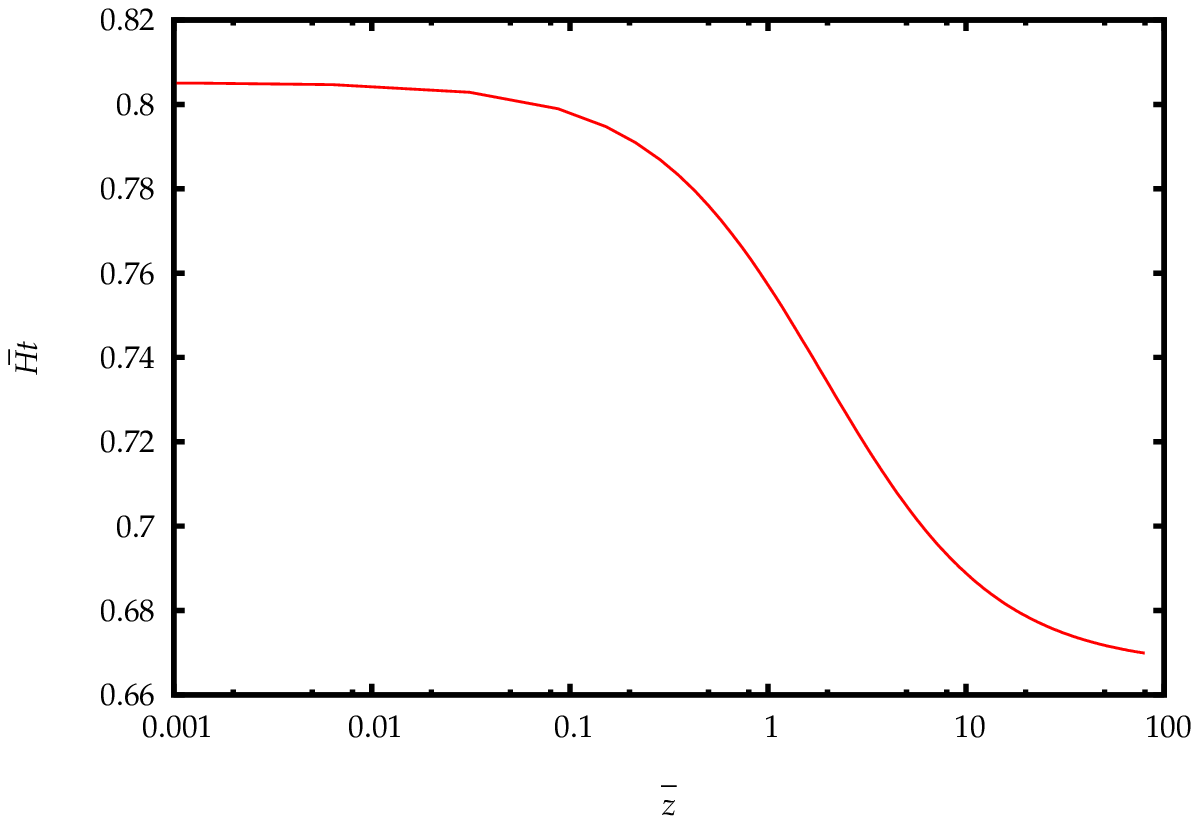}}
\begin{center} {\bf (a)} \end{center}
\end{minipage}
%\hfill
\begin{minipage}[t]{7.7cm}
\scalebox{1.0}{\includegraphics[angle=0, clip=true, trim=0cm 0cm 0cm 0cm, width=\textwidth]{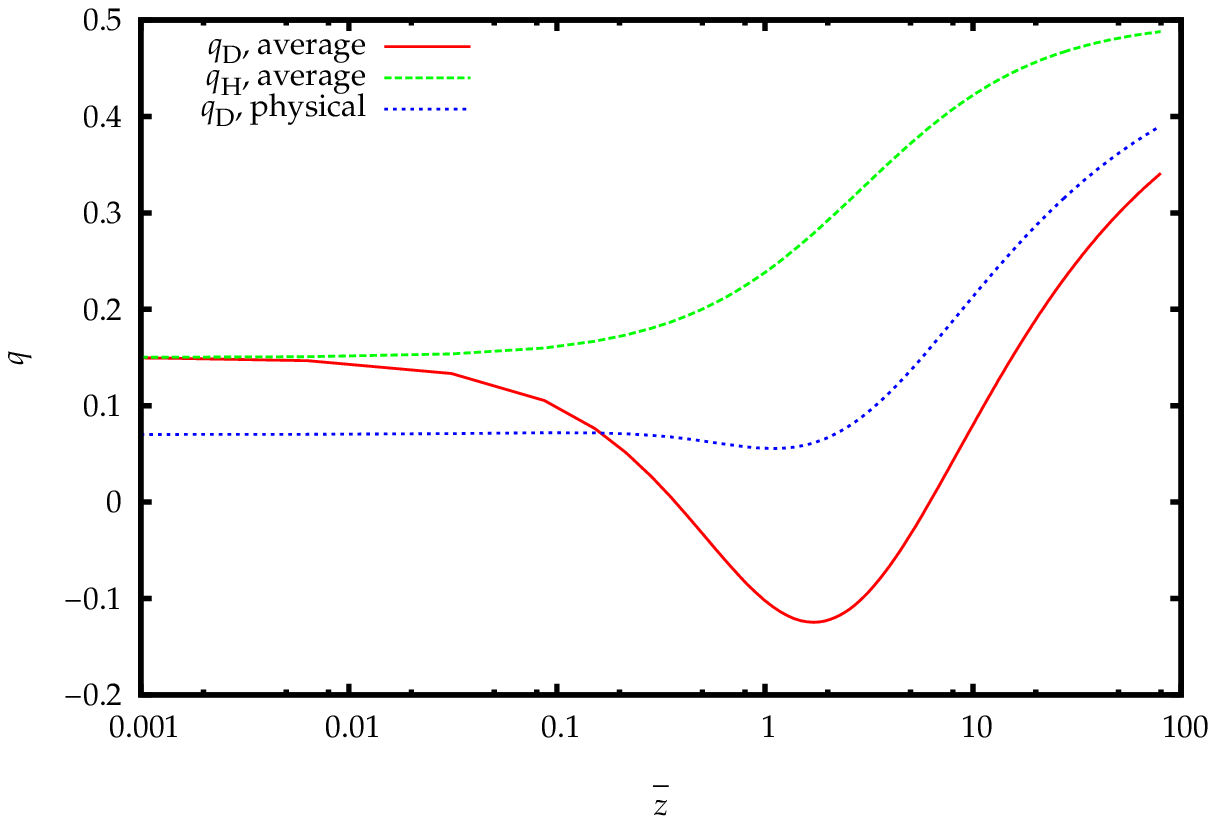}}
\begin{center} {\bf (b)} \end{center}
\end{minipage}
%\hfill
\caption{a) The average expansion rate times the age of the universe, $\bH t$,
as a function of the mean redshift \re{bz}.
b) The deceleration parameters $q_H$ and $q_D$ as a function of the
the mean redshift \re{bz}. The quantity $q_D$ corresponding to the
physical distance is shown as a function of the physical redshift instead;
see \sec{sec:sig}.}
\label{fig:exp}
\end{figure}

\begin{wrapfigure}[16]{r}[0pt]{0.5\textwidth}
%\begin{figure}
\scalebox{0.5}{\includegraphics[angle=0, clip=true, trim=0cm 0cm 0cm 0cm, width=1.0\textwidth]{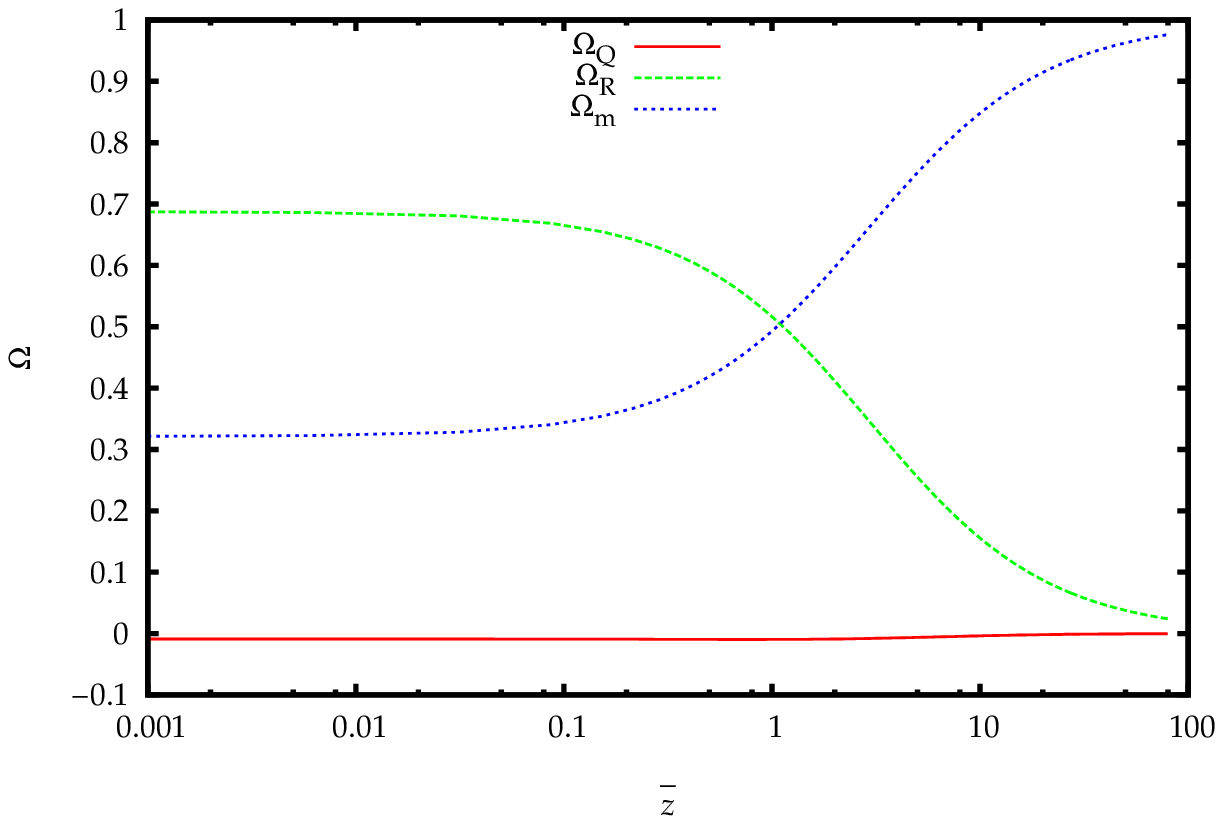}}
\caption{The density parameters $\Om$, $\OR$ and $\OQ$
as a function of the mean redshift \re{bz}.}
\label{fig:Omegas}
%\end{figure}
\end{wrapfigure}

The average expansion rate times the age of the universe, $\bH t$,
is shown in \fig{fig:exp}a as a function of the mean redshift $\bz$
(defined below in \re{bz}).
At large redshifts (corresponding to early times),
the average expansion rate is close to the EdS case, but
at redshifts of order unity and smaller the holes
expand significantly faster than the background,
so $\bH t$ grows. In a universe completely dominated
by completely empty voids, we would have $\bH t=1$.
Note that the average expansion rate only decelerates less,
it does not accelerate. The corresponding deceleration parameter
$q_H$ is shown in \fig{fig:exp}b, and it is positive at all times.
(We also show the deceleration parameters $q_D$ that correspond
to the distance instead of the average expansion rate; see \sec{sec:sig}.)
The density parameters are shown in \fig{fig:Omegas}.
The values today are $\Omn=0.32, \ORn=0.69$ and $\OQn=-0.01$.
The backreaction variable is small at all times, $|\OQ|\lesssim0.01$.
This does not mean that inhomogeneities have a small effect on the
expansion rate, just that the effect reduces to a spatial curvature term
that evolves like $\propto \ba^{-2}$, \ie in the same way as in the FRW
case \cite{Buchert:1999b, Rasanen:2005, Rasanen:2008a}.

\para{Mean redshift and angular diameter distance.}

In \cite{Rasanen:2008b, Rasanen:2009b} it was argued that
in a statistically homogeneous and isotropic dust
universe in which the distribution evolves slowly,
the redshift is approximately equal to the mean redshift
over distances larger than the homogeneity scale,
\bea \label{bz}
  1 + \bz \equiv \ba(t)^{-1} = e^{ \int_t^{t_0}\rmd t' \bH(t') } \ ,
\eea

\noindent and the angular diameter distance can to first
approximation be solved from
\bea \label{DAH}
  \bH \frac{\rmd}{\rmd\bz} \left[ (1+\bz)^2 \bH \frac{\rmd D_A}{\rmd\bz} \right] &=& - 4\pi\GN \av{\rhom} D_A \ ,
\eea

\noindent where $4\pi\GN\av{\rhom}=\frac{3}{2}\Omn \bH_0^2 (1+\bz)^3$ due
to \re{cons} and \re{bz}. The equation \re{DAH} determines the
distance, given $\bH(\bz)$ and $\Omn$.
The luminosity distance is $D_L(\bz)=(1+\bz)^2 D_A(\bz)$
\cite{Etherington:1933, Ellis:1971}. The quantities calculated
from \re{bz} and \re{DAH} are mean values, with small variations
expected for typical light rays, and possibly large variations for
exceptional lines of sight, e.g.\ in the case of strong lensing.

\subsection{Averages and light propagation} \label{sec:avelight}

\begin{figure}
%\hfill
\begin{minipage}[t]{7.7cm} 
\scalebox{1.0}{\includegraphics[angle=0, clip=true, trim=0cm 0cm 0cm 0cm, width=\textwidth]{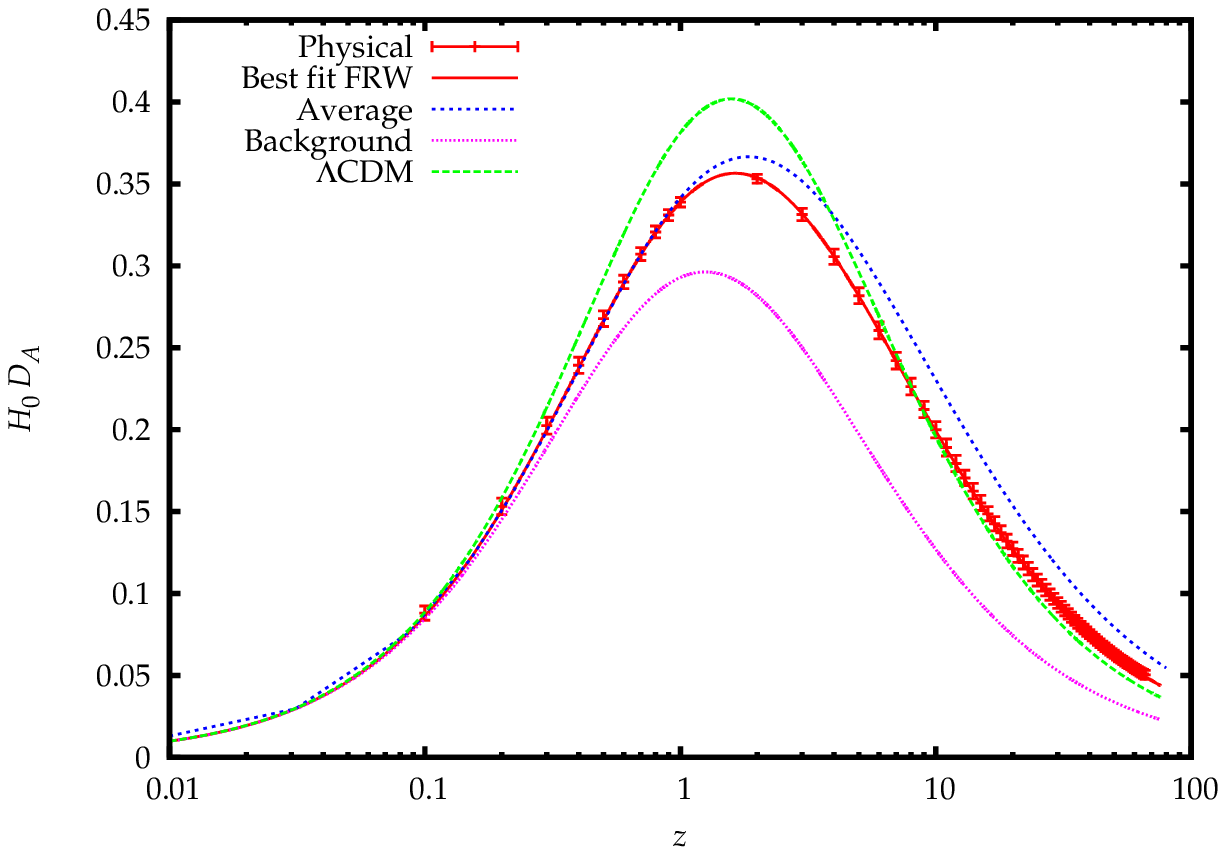}}
\begin{center} {\bf (a)} \end{center}
\end{minipage}
%\hfill
\begin{minipage}[t]{7.7cm}
\scalebox{1.0}{\includegraphics[angle=0, clip=true, trim=0cm 0cm 0cm 0cm, width=\textwidth]{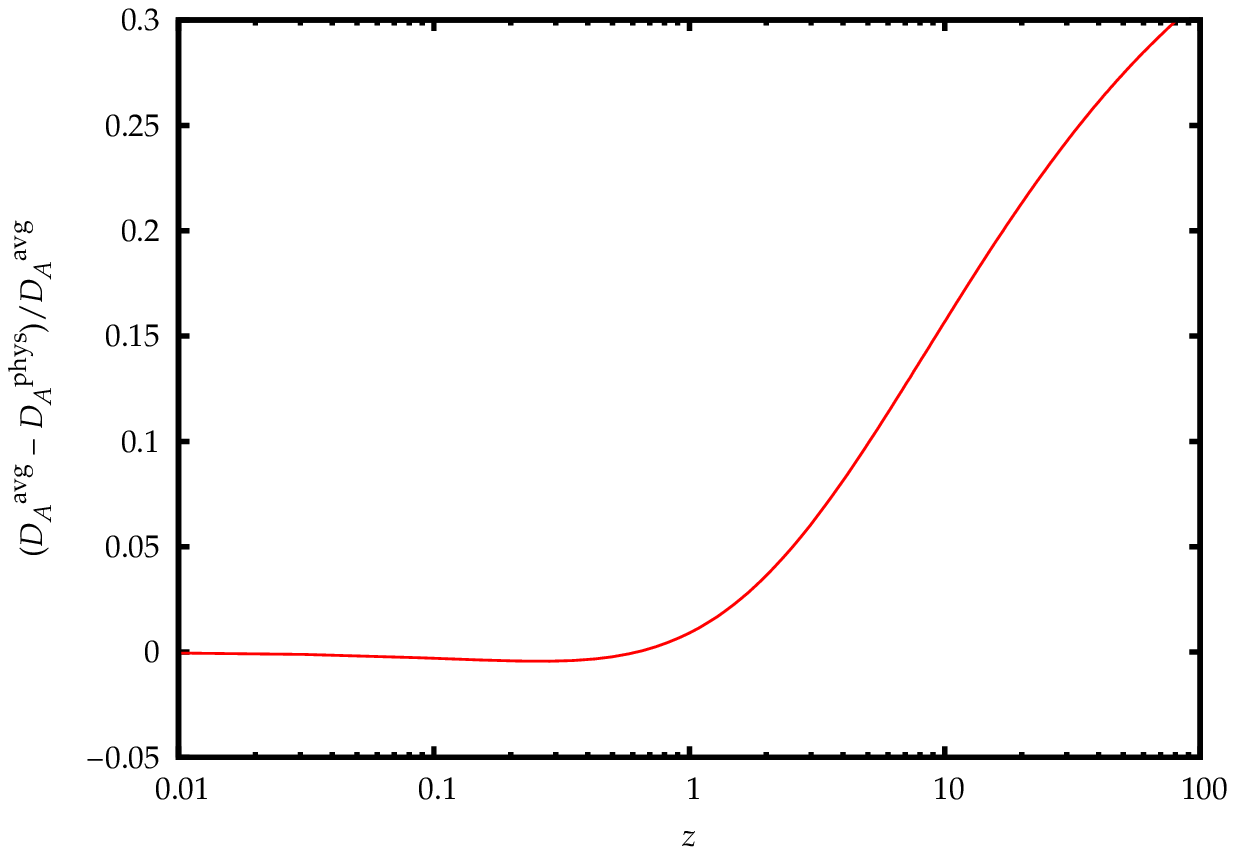}}
\begin{center} {\bf (b)} \end{center}
\end{minipage}
%\hfill
\caption{a) The angular diameter distance calculated from the
light propagation equations, from the average expansion rate
and in the background model. For the physical distance the
error bars show standard deviation for 1000 light rays, and the line
shows an FRW model that gives a good fit. Distance in the spatially
flat \LCDM model with the same $\Omn=0.32$ as in the
physical model is also shown for comparison.
b) The relative difference of the angular diameter distance
corresponding to the best-fit FRW model and the angular
diameter distance calculated from the average expansion rate.
}
\label{fig:DA}
\end{figure}

\para{Angular diameter distance.}

Let us now see how well the mean quantities calculated
from \re{bz} and \re{DAH} describe the redshift and the angular
diameter distance calculated with the light propagation
equations, as discussed in \sec{sec:light}. In \fig{fig:DA} we compare
the distance calculated from the null geodesic equation and the
Sachs equations as a function of the physical redshift,
the mean distance calculated from from the average expansion
rate using \re{DAH} as a function of the mean redshift \re{bz},
and the background distance as a function of the background
redshift \mbox{$1+\zb=a(t)^{-1}$}. The distance-redshift relation
of the \LCDM model with the same $\Omn=0.32$ as the Swiss Cheese
model is also shown for comparison.

The physical distance\footnote{Recall that what we call the
``physical distance'' is based on the continuous solution
of the Sachs equations \re{thetaeq} and \re{sigmaeq},
with the jumps due to the surface layers neglected.}
is very different from the background distance.
It is also slightly different for different light rays.
We have propagated 1000 light rays and fit a FRW model
with dust, spatial curvature and vacuum energy to
the resulting points. A model with
$\Omn=0.42$, $\Omega_{K0}=0.44$ and $\Omega_{\Lambda 0}=0.14$
gives an excellent description of the physical distance.
The distance calculated from the average expansion rate also gives
a reasonable description of the physical distance up to $z\approx1$,
but for higher redshifts the agreement is rather poor. At
$z=100$ the distance calculated from the average expansion rate
overestimates the real distance by more than 30\%.
The discrepancy is related to the fact that the integrated
expansion rate and density along a light ray are different
from the average quantities. Looking at the redshift will show
this in more detail.

\para{Redshift.}

\begin{figure}
%\hfill
\begin{minipage}[t]{7.7cm} 
\scalebox{1.0}{\includegraphics[angle=0, clip=true, trim=0cm 0cm 0cm 0cm, width=\textwidth]{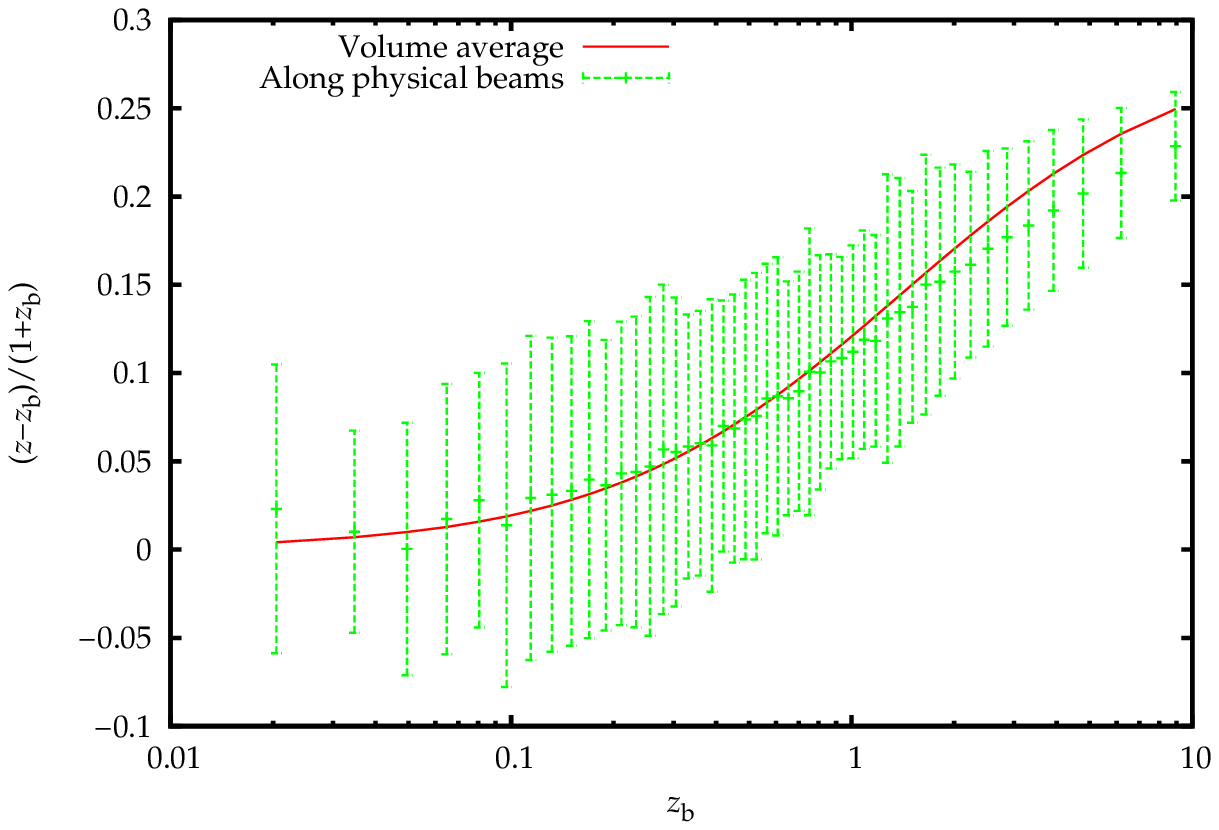}}
\begin{center} {\bf (a)} \end{center}
\hfill
\end{minipage}
%\hfill
\begin{minipage}[t]{7.7cm}
\scalebox{1.0}{\includegraphics[angle=0, clip=true, trim=0cm 0cm 0cm 0cm, width=\textwidth]{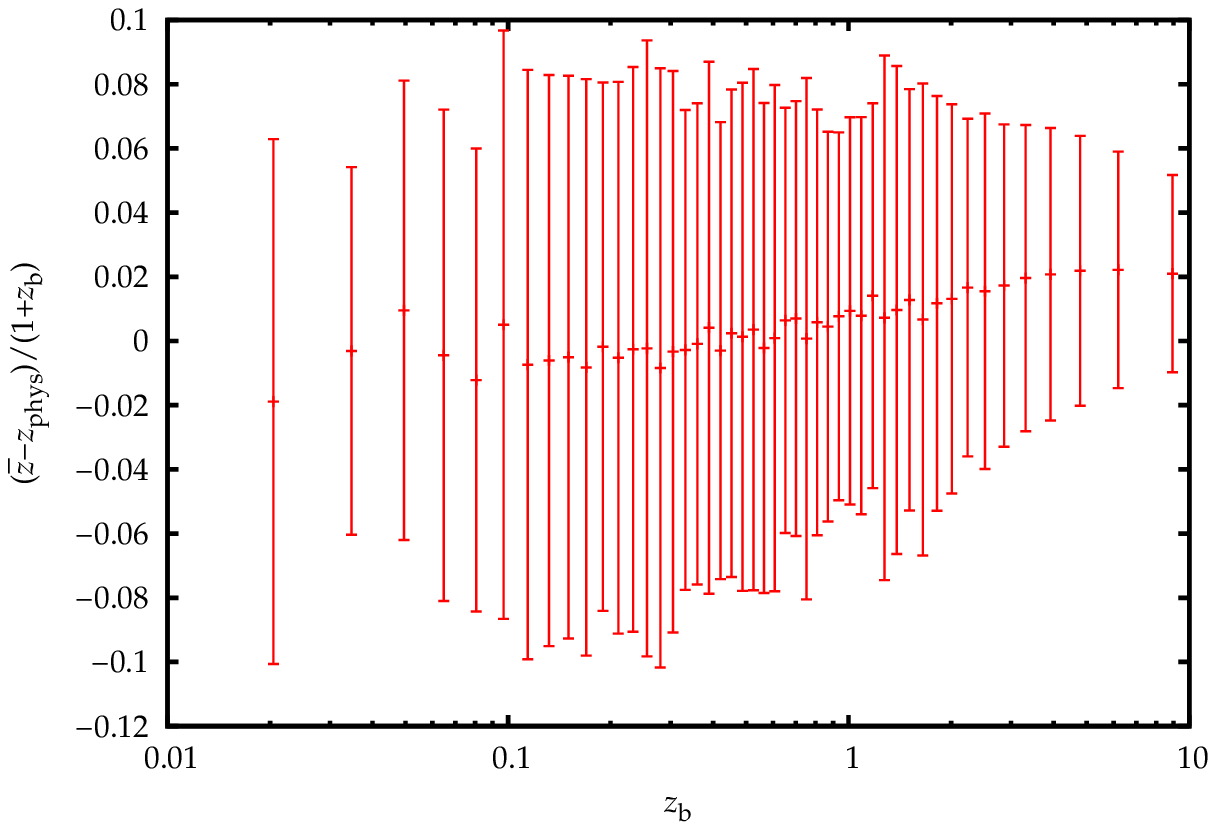}}
\begin{center} {\bf (b)} \end{center}
\hfill
\end{minipage}
%\hfill
\begin{minipage}[t]{7.7cm}
\scalebox{1.0}{\includegraphics[angle=0, clip=true, trim=0cm 0cm 0cm 0cm, width=\textwidth]{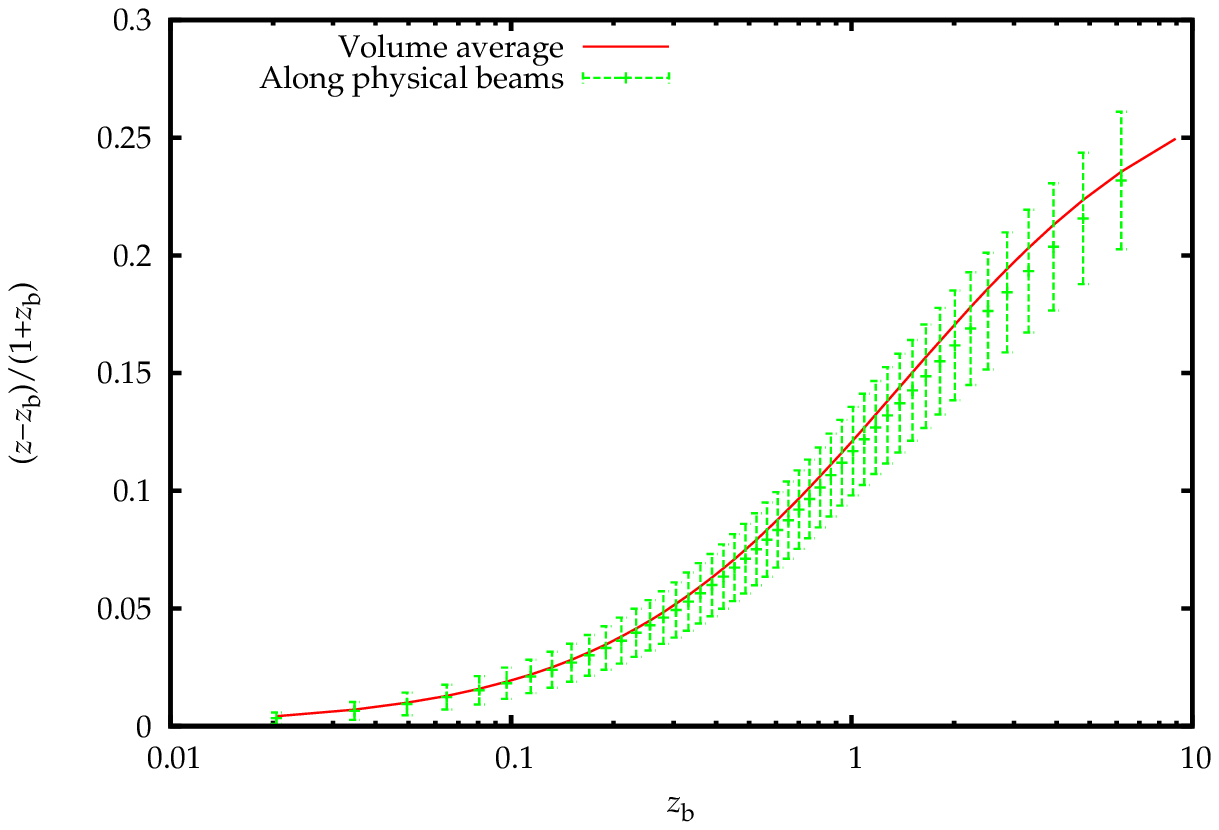}}
\begin{center} {\bf (c)} \end{center}
\end{minipage}
%\hfill
\begin{minipage}[t]{7.7cm}
\scalebox{1.0}{\includegraphics[angle=0, clip=true, trim=0cm 0cm 0cm 0cm, width=\textwidth]{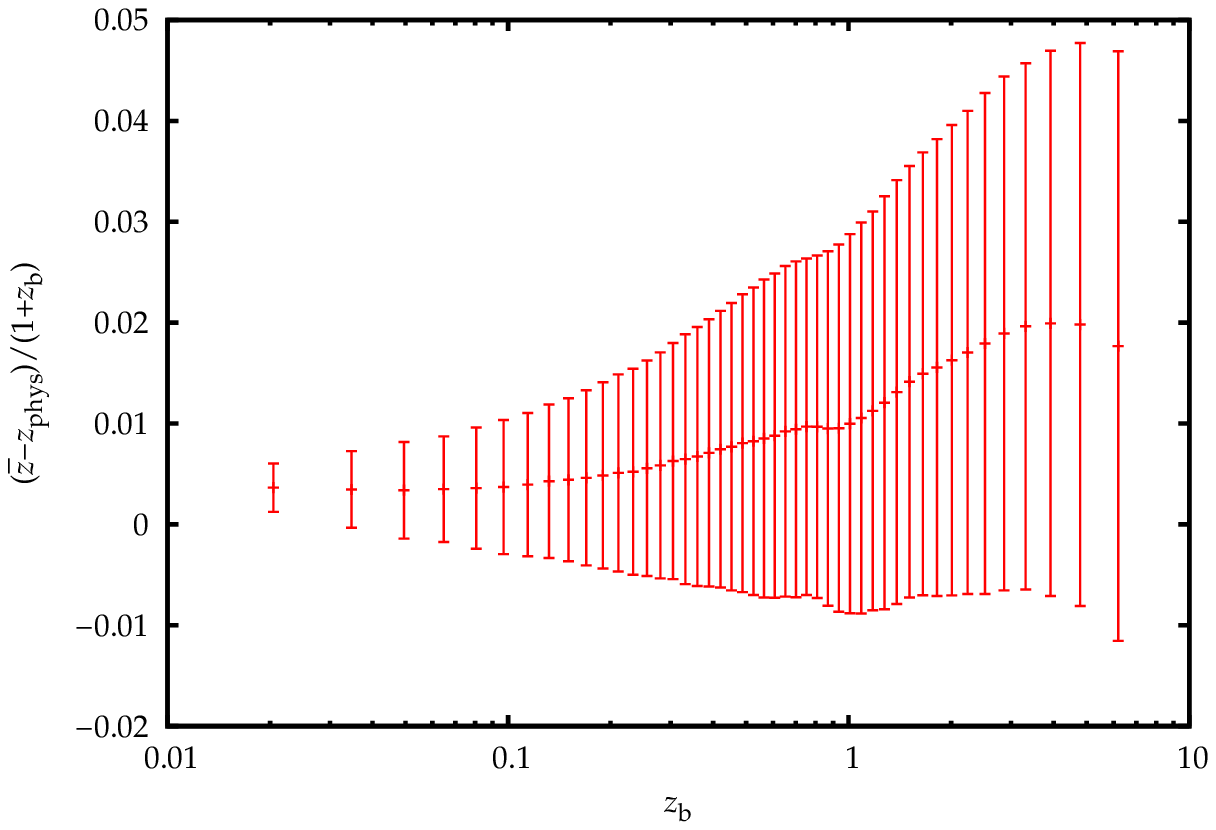}}
\begin{center} {\bf (d)} \end{center}
\end{minipage}
%\hfill
\caption{a) The relative difference to the background redshift
for the redshift calculated from the null geodesic equation
and the mean redshift calculated from the average expansion
rate. The error bars show the standard deviation for 1000 light rays.
b) The relative difference between the redshift calculated
from the null geodesic equation and the mean redshift.
c) The same as a), but with the redshift calculated along null
curves that are straight relative to the background instead
of using the null geodesic equation.
d) The same as b), but with the redshift calculated along null
curves that are straight relative to the background instead
of using the null geodesic equation.
}
\label{fig:z}
\end{figure}

We consider the redshift calculated from the null geodesic
equations without change, as well as a modified redshift
calculated from null curves that are straight according
to the background metric.
In \fig{fig:z}a we show the relative deviation from the background
for the redshift calculated from the null geodesic equation as well
as for the mean redshift $\bz$.
In \fig{fig:z}b we show the relative difference between the
physical redshift and the mean redshift.
As with the distance, the physical redshift is quite
different from the background redshift, and the difference
grows monotonically.
The reason is that the expansion rate (and the shear) along the
light ray are different from the background expansion rate.
The mean redshift \re{bz} calculated from the average expansion
rate has the same systematic evolution as the physical redshift,
with statistical deviations of order 10\% for small redshifts
and 5\% for high redshifts.
There appears to be a small systematic offset at large redshifts,
though it should be noted that as the bins are equally spaced
in time, the largest redshifts are sparsely sampled.
In figures \ref{fig:z}c and \ref{fig:z}d we show the same
comparison for the redshift calculated along null curves
that are straight according to the background metric, \ie
with the sharp turns due to the surface layers neglected.
In this case, the mean redshift agrees with the physical redshift
even better, with a statistical error of less than 5\% at all redshifts,
showing that most of the discrepancy between the physical redshift
and the mean redshift in the full case is due to the surface layers.

\begin{figure}
%\hfill
\begin{minipage}[t]{7.7cm} 
\scalebox{1.0}{\includegraphics[angle=0, clip=true, trim=0cm 0cm 0cm 0cm, width=\textwidth]{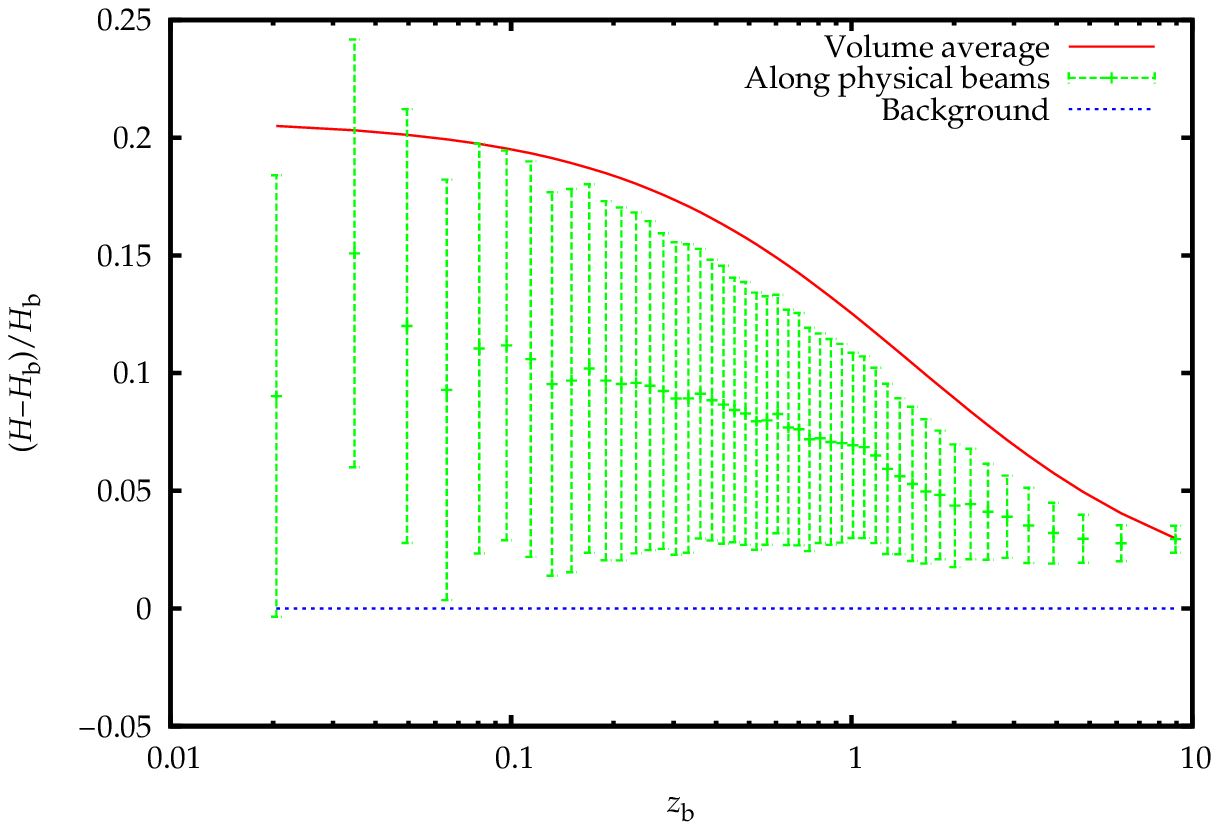}}
\begin{center} {\bf (a)} \end{center}
\end{minipage}
%\hfill
\begin{minipage}[t]{7.7cm}
\scalebox{1.0}{\includegraphics[angle=0, clip=true, trim=0cm 0cm 0cm 0cm, width=\textwidth]{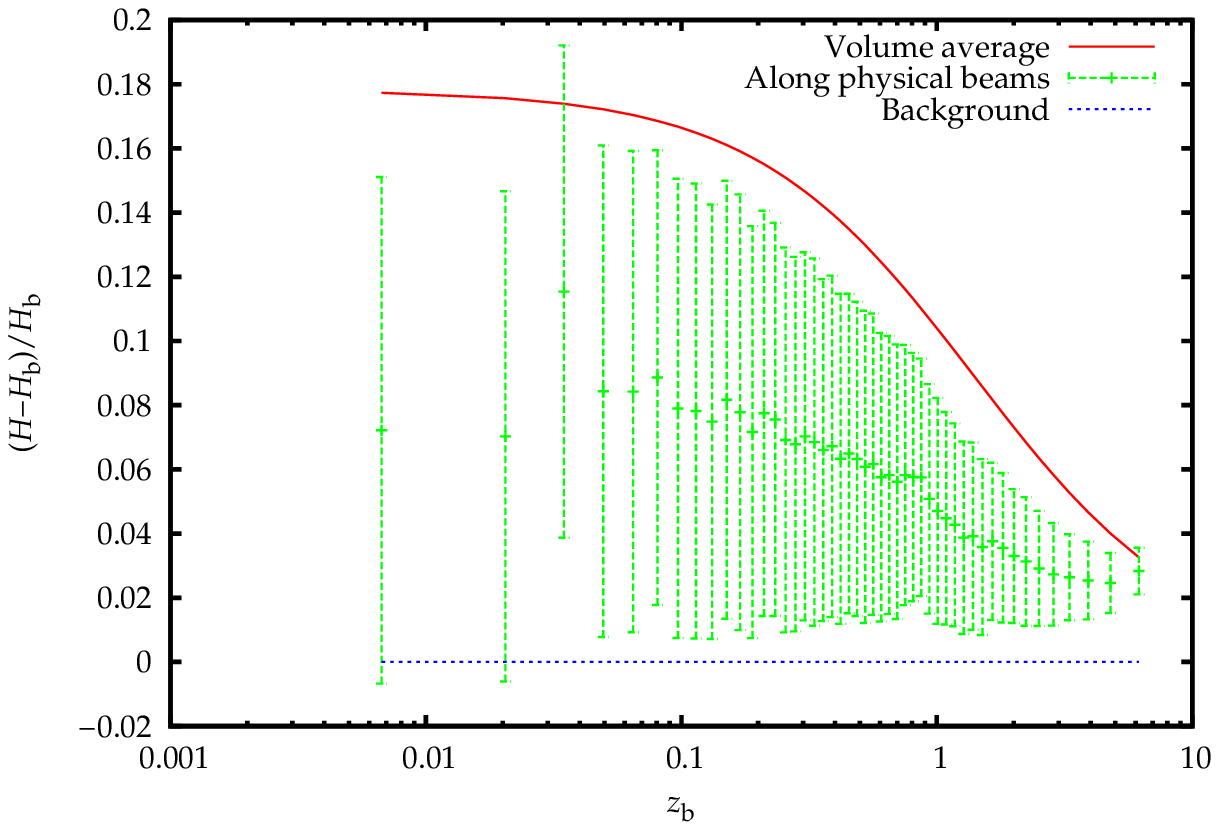}}
\begin{center} {\bf (b)} \end{center}
\end{minipage}
%\hfill
\caption{a) The relative difference from the background expansion
rate for the expansion rate along the light ray and the
the spatial average of the expansion rate.
The error bars show the standard deviation for 1000 light rays.
b) The same plot for the case when the light rays go straight
according to the background metric.}
\label{fig:theta}
\end{figure}

It is notable that in neither case is the mean expansion rate
along the light ray the same as the spatial average.
In \fig{fig:theta} we show the relative deviation from the
background expansion rate for the expansion rate along the light ray
and the spatial average of the expansion rate.
For both treatments of the physical redshift, the expansion rate
seen by the light rays is between the background value and the
spatial average. This is also true for the energy density.

\begin{figure}
%\hfill
\begin{minipage}[t]{7.7cm} 
\scalebox{1.0}{\includegraphics[angle=0, clip=true, trim=0cm 0cm 0cm 0cm, width=\textwidth]{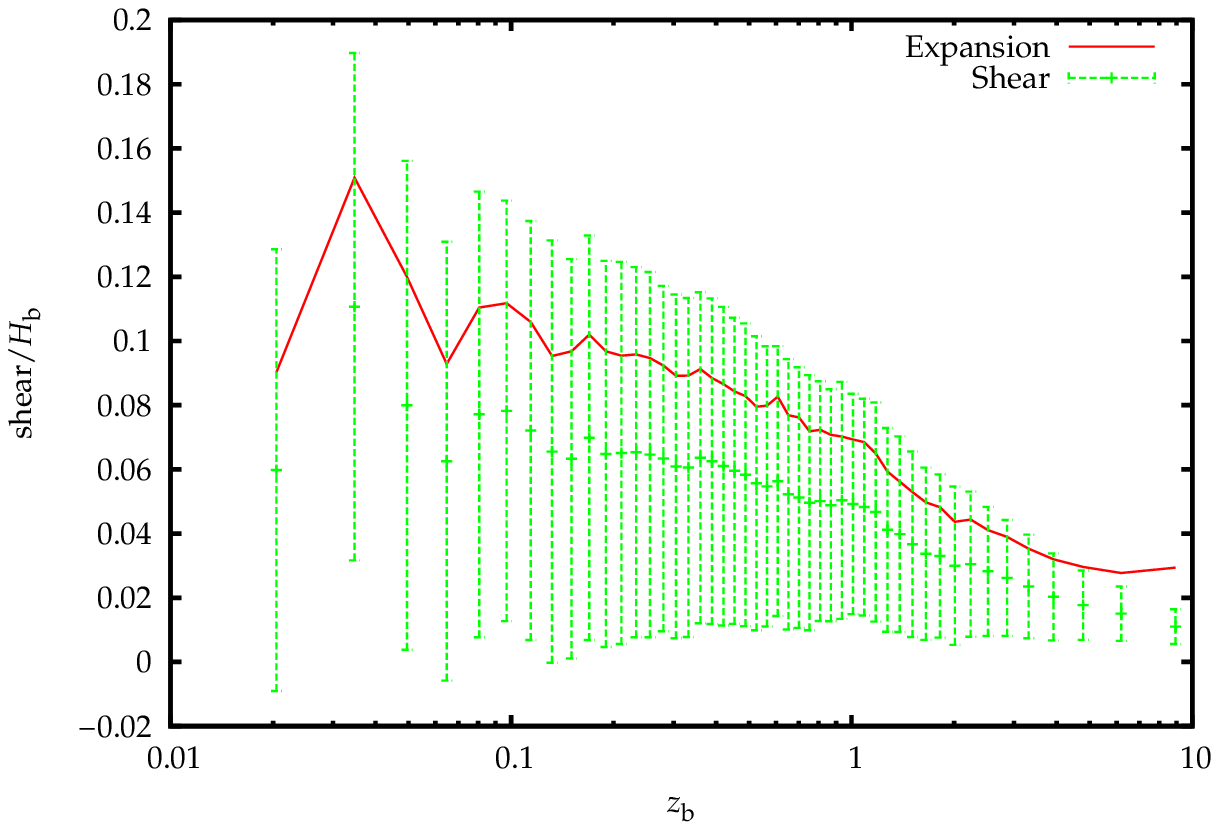}}
\begin{center} {\bf (a)} \end{center}
\end{minipage}
%\hfill
\begin{minipage}[t]{7.7cm}
\scalebox{1.0}{\includegraphics[angle=0, clip=true, trim=0cm 0cm 0cm 0cm, width=\textwidth]{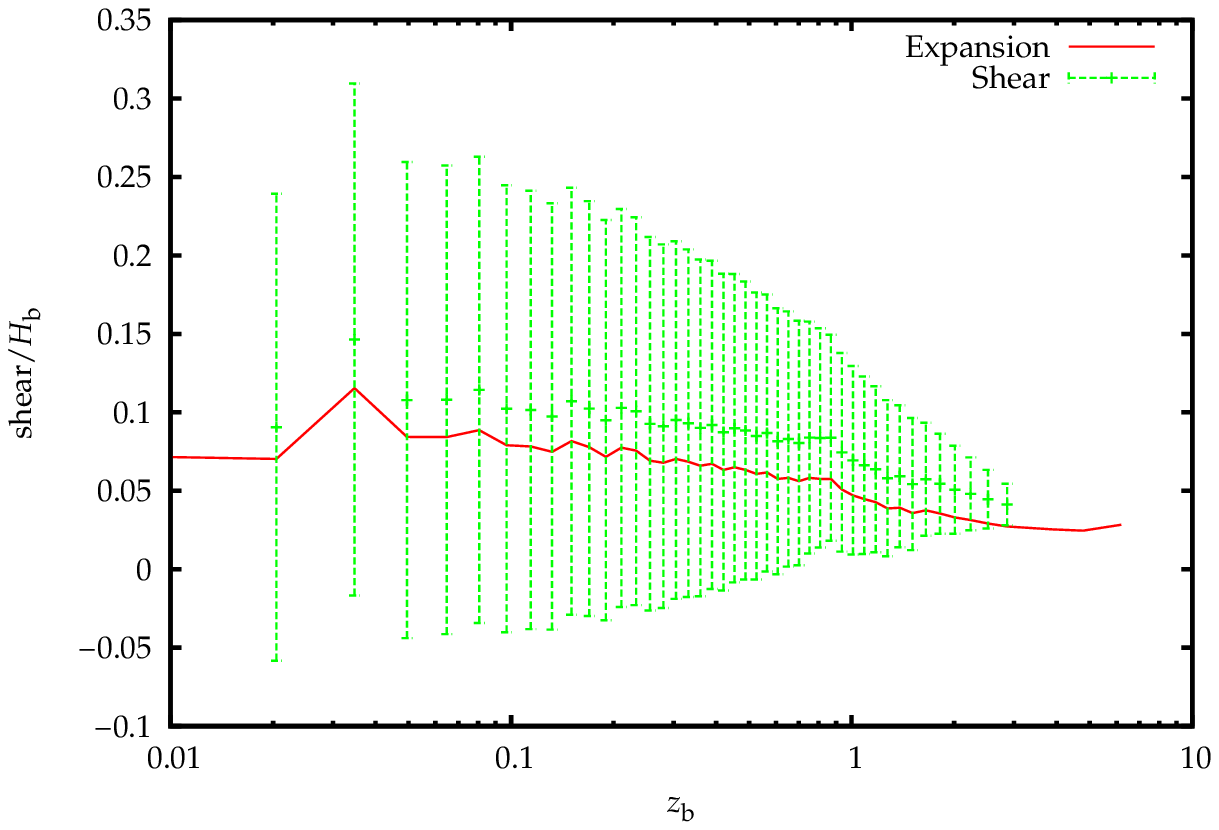}}
\begin{center} {\bf (b)} \end{center}
\end{minipage}
%\hfill
\caption{a) The shear normalised to the background expansion rate,
$\sigma_{\a\b}e^\a e^\b/H$, along the null geodesic.
The error bars show the standard deviation for 1000 light rays.
The red line shows the relative deviation of the expansion
rate along the line of sight from the background expansion rate
(\ie the quantity plotted in \fig{fig:z}), without error bars.
b) The same plot for the case when the light rays go straight
according to the background metric.}
\label{fig:zshear}
\end{figure}

In \cite{Rasanen:2008b, Rasanen:2009b} it was argued that the
mean expansion rate along a light ray and the spatial average
of the expansion rate should be close to each other for slowly
evolving and small structures with a statistically homogeneous and
isotropic distribution. The argument was that in the integral
\re{zcov}, one would the contributions of
$\frac{1}{3}\Delta\theta\equiv\frac{1}{3}\left(\theta-\av{\theta}\right)$
and $\sigma_{\a\b}e^\a e^\b$ to be highly suppressed for symmetry
reasons. In the present case, these two terms cancel each
other, as in the near-FRW case \cite{Rasanen:2011b}, but they do
not cancel individually. The expansion rate seen by a typical light
ray is quite different from the spatially averaged expansion rate,
but this difference cancels with the shear to
leave only the contribution of the average expansion rate.
The importance of the shear can be seen in \fig{fig:zshear},
which shows that the shear is comparable to
the deviation of the expansion rate along the null geodesic
from the background expansion rate. This is unexpected, and it
would be interesting to repeat the analysis in an exact
solution that does not have surface layers.

\subsection{Signatures of backreaction} \label{sec:sig}

\para{FRW consistency parameters.}

If backreaction is significant, the relation between the average
expansion rate and the distance is in general different from the FRW
case. This is an important
aspect of backreaction, which cannot be reproduced by any model
based on the four-dimensional FRW metric, regardless of the matter
content or the equation of motion \cite{Rasanen:2008b, Boehm:2013}.

In addition to the average expansion rate $\bH$ that describes
how the volume of the universe evolves, it is useful to define a
'fitting' expansion rate $\Hfit$ that describes how the distance
evolves \cite{Boehm:2013}.
Specifically, $\Hfit(z)$ is the expansion rate of the spatially flat FRW
``fitting model'' that has the same distance-redshift relation $D_A(z)$
as the backreaction model.
In a FRW model, the angular diameter distance is\footnote{Assuming that
the expansion rate is strictly monotonic, so that redshift is a valid time
coordinate, and non-negative.}
\bea \label{DAFRW}
  D_A(z) = (1+z)^{-1} \frac{1}{\sqrt{-K}} \sinh\left(\sqrt{-K}\int_0^z \frac{\rmd \tilde z}{H(\tilde z)} \right) \ ,
\eea

\noindent where $K$ is the spatial curvature constant defined in
\re{FRW}. We thus have, defining $D\equiv (1+z) D_A$,
\bea
  \Hfit(z) = \frac{1}{D'(z)} \ ,
\eea

\noindent where the prime refers to derivative with respect to
$z$, not $r$. As the distance is normalised to
$D\simeq \bH_0^{-1} z$ for $z\ll1$, the expansion rates
$\Hfit$ and $\bH$ agree at small redshift, and
(if backreaction is significant) disagree at large redshift.

As in \cite{Clarkson:2007b}, we can solve $K$ from \re{DAFRW} to obtain
\bea \label{K}
  K &=& \frac{ 1 - ( H D' )^2 }{ D^2 } \ .
\eea

\noindent The relation \re{K} is now taken as the definition of a
function $K(z)$ in the general case when the universe is not
necessarily described by a FRW model, with the FRW expansion rate
$H$ replaced by the average expansion rate $\bH$.
(In general, $K$ would also be function of angular direction,
but for distances larger than the homogeneity scale the directional
dependence is expected to be small if the universe is statistically
homogeneous and isotropic \cite{Rasanen:2008b, Rasanen:2009b}.)
In terms of the fitting model expansion rate, we have
\bea \label{Kz}
  K(z) &=& \frac{ 1 - \left[ \bH(z)/\Hfit(z) \right]^2 }{ \left[ \int_0^z \rmd \tilde z/\Hfit(\tilde z) \right]^2 } \ .
\eea

\noindent In \cite{Clarkson:2007b}, the following quantity was also introduced
to quantify the deviation of $K$ from constant:
\bea \label{C}
  \mC \equiv - \frac{D^3}{2 D'} K' = 1 + \bH^2 ( D D'' - D'^2  ) + \bH \bH' D D' \ .
\eea

If the average expansion rate, redshift and distance are well
described by the FRW model, $K$ is constant and equal to $-\bH_0^2\ORn$. 
Observational ranges for $K$ were determined in
\cite{Shafieloo:2009, Mortsell:2011}, with the result $|K|/\bH_0^2\lesssim1$,
though there may be significant systematic uncertainty.
If $K$ is observed to vary with redshift, the universe cannot
be described by any four-dimensional FRW model.
On the other hand, a magnitude of $0.1\lesssim|K|/\bH_0^2\lesssim1$
is expected if backreaction is significant in the real universe and
light propagation is well described by \re{bz} and \re{DAH} \cite{Boehm:2013},
which is the case in our model for redshifts below unity.
However, a constant $K$ does not rule out significant backreaction,
because clumpiness could in principle change the expansion rate
and distance in a way that preserves the FRW relation.
In particular, if the average expansion rate and distance
are related by \re{DAH} and clumpiness changes the average expansion
rate by $\bH(z)^2\rightarrow \bH(z)^2 + A + B (1+z)^2 + C (1+z)^3$, where
$A$, $B$ and $C$ are constants (\ie like a mixture of dust,
spatial curvature and vacuum energy in the FRW model), the relation
between the expansion rate and distance is identical to the FRW case.
Likewise, $K'\neq0$ does not necessarily indicate backreaction,
as the relation between the expansion rate and the distance is also
altered in models in which the universe is spherically symmetric (and
inhomogeneous) on Gpc scales \cite{February:2009} and in some models
with extra dimensions \cite{Ferrer}.

\begin{figure}[t]
\hfill
\begin{minipage}[t]{4.8cm}
\scalebox{1.0}{\includegraphics[angle=0, clip=true, trim=0cm 0cm 0cm 0cm, width=\textwidth]{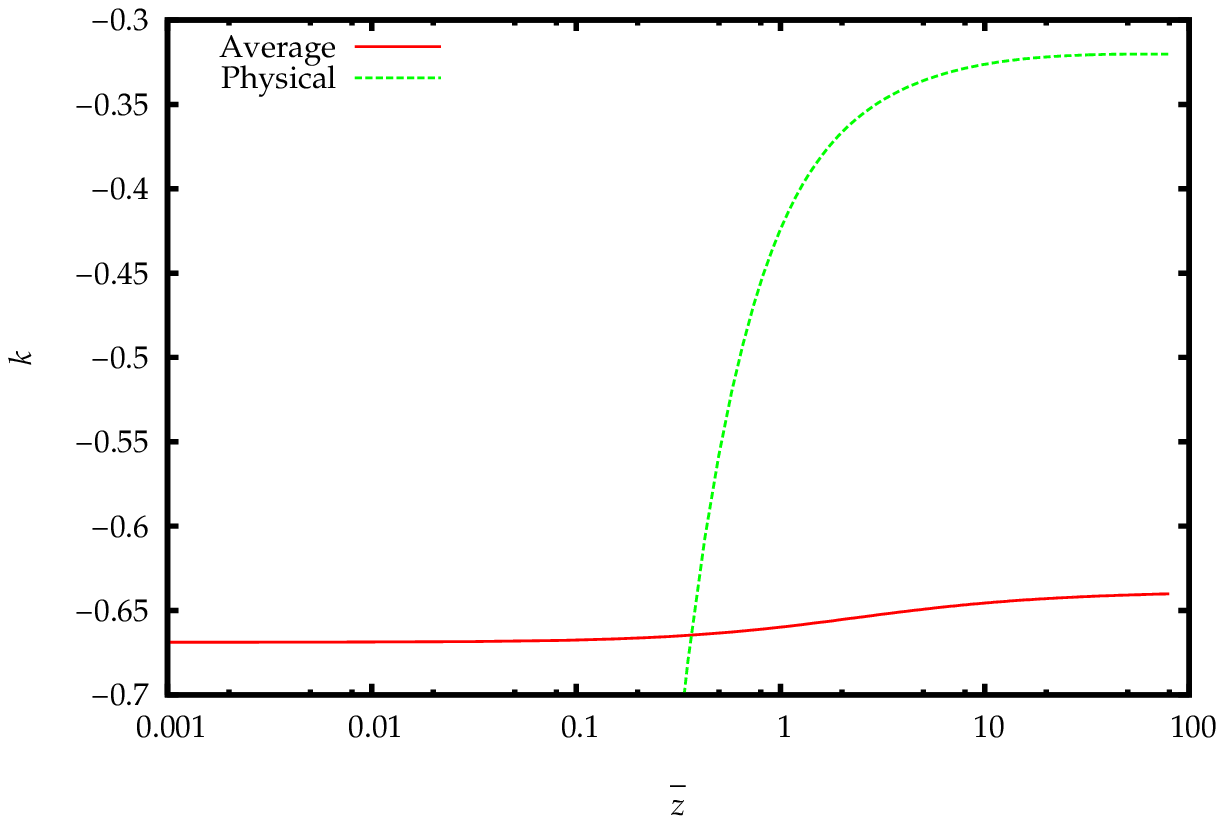}}
\begin{center} {\bf (a)} \end{center}
\end{minipage}
%\hfill
\begin{minipage}[t]{4.8cm}
\scalebox{1.0}{\includegraphics[angle=0, clip=true, trim=0cm 0cm 0cm 0cm, width=\textwidth]{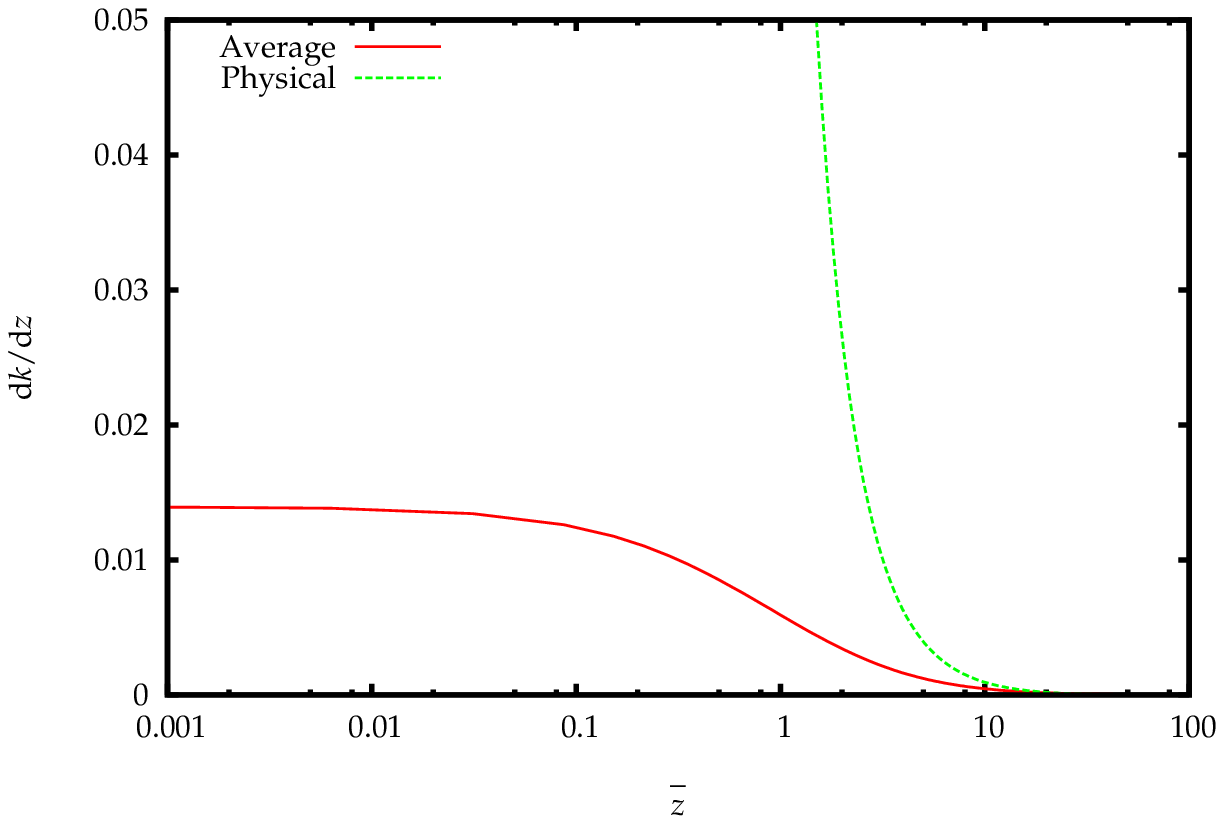}}
\begin{center} {\bf (b)} \end{center}
\end{minipage}
%\hfill
\begin{minipage}[t]{4.8cm}
\scalebox{1.0}{\includegraphics[angle=0, clip=true, trim=0cm 0cm 0cm 0cm, width=\textwidth]{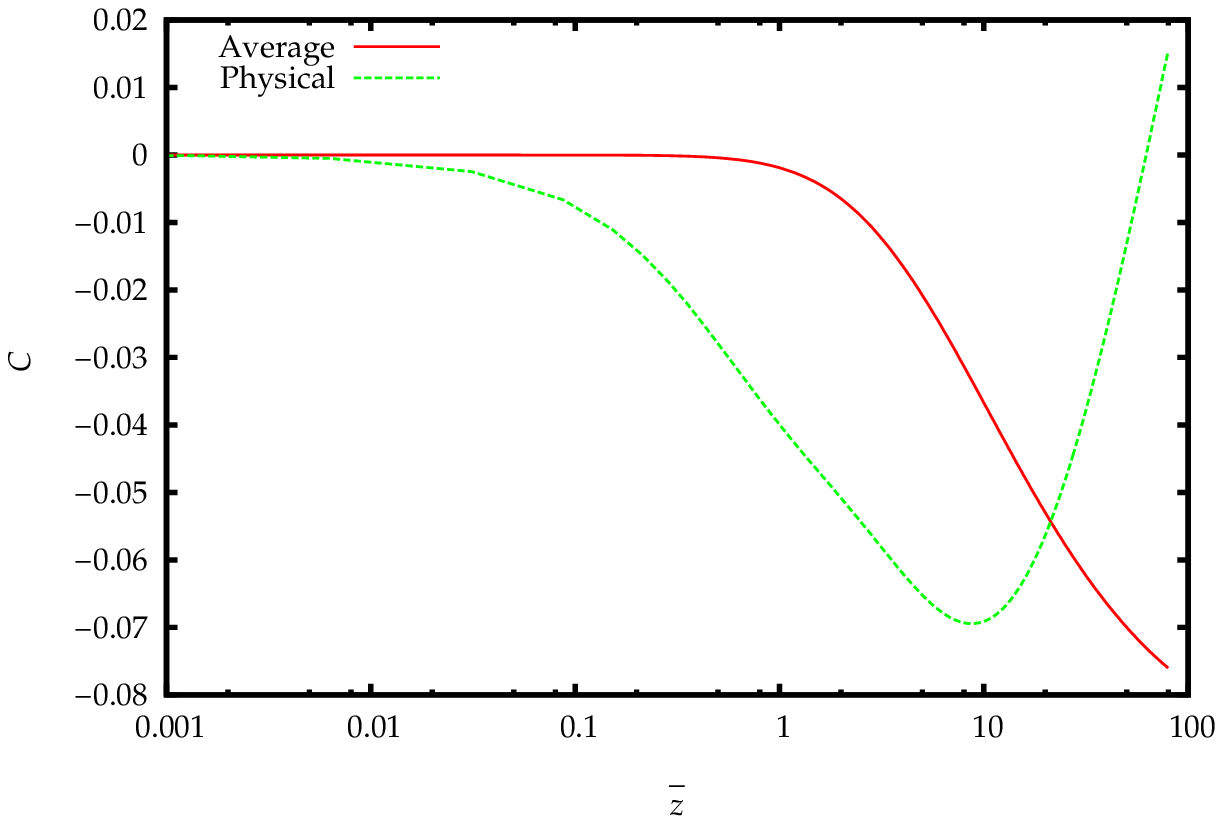}}
\begin{center} {\bf (c)} \end{center}
\end{minipage}
%\hfill
\caption{a) The FRW consistency parameter $k\equiv K/\bH_0^2$.
b) The $z$-derivative of $k$.
c) The FRW consistency parameter $\mC$.}
\label{fig:K}
\end{figure}

In \fig{fig:K} we show $K$, $K'$ and $\mC$ both for the distance
calculated from the average expansion rate using \re{bz} and \re{DAH}
and for the physical distance (more precisely, we use the FRW model
that gives an excellent fit to the physical distance calculated
from 1000 light rays, as discussed in \sec{sec:avelight}).
For the average expansion rate, the quantity $K$ is almost constant,
and $K'$ and $\mC$ are small. This corresponds to the fact that
backreaction in the model changes the average expansion rate
in almost the same way as FRW negative spatial curvature.
The behaviour of $K$, $K'$ and $\mC$ is very different in the case
of the physical distance. For large redshifts this is to be expected,
because the two distances agree well only at $z\lesssim1$.
The discrepancy at small $z$, in turn, is related to the fact that the
denominator of $K$ vanishes like $z^2$ at small $z$.
For the distance solved from \re{DAH}, the numerator goes to zero
at the same rate, so $K$ is finite at $z=0$.
However, for the physical distance the numerator only vanishes like $z$,
so $K$ diverges as $1/z$ as $z$ approaches zero.
From the observational point of view, the fact that an arbitrarily small
mismatch between $D'$ and $\bH$ in \re{Kz} leads to divergent $K$ and $K'$
means that the error bars on $K$ and $K'$ diverge at small $z$.
The consistency parameter $\mC$ is defined in such a way that it
does not diverge at $z=0$, so it does not suffer from this problem.
The quantities $\mC$ calculated from the two distances both go to zero
as z goes to zero, but they differ markedly already at $z=0.1$, even though
the distances agree well, as shown in \fig{fig:DA}. The reason
is that $\mC$ is sensitive to the second derivative of the distance,
which can be quite different even when the distances are close.

\para{Effective equations of state.}

Another useful way of looking at the deviation of the
distance--expansion rate relation from the FRW case is to
define separate effective equations of state for the
expansion rate and the distance \cite{Rasanen:2008b, Boehm:2013}.
The effective expansion rate equation of state $w_H$ is defined as
the equation of state of an extra energy density component
that would give the expansion history $\bH(z)$ in a spatially flat
FRW model (with the same value of the dust density parameter $\Omn$
as in the backreaction model). From \re{Ray}, \re{Ham} and \re{q} we have
\bea \label{wH}
  w_H &=& \frac{ 2 (1+z) \bH \bH' - 3 \bH^2 }{3 \bH^2 - 8\pi\GN \av{\rhom}} = \frac{2 q_H - 1}{3 - 3 \Om} \ .
\eea

\noindent This definition is somewhat inconvenient, because
$w_H$ diverges if at some moment the expansion rate is the
same as in (and the acceleration is different from) the EdS model.
In particular, this happens if the expansion rate decelerates first
less and then more than in the EdS model, as could be expected if
the expansion eventually accelerates
\cite{Rasanen:2006a, Rasanen:2006b, Boehm:2013}.
(This also happens in some models with extra dimensions \cite{Ferrer}.)
It is more informative to consider the total equation of state defined as
\bea \label{wHtot}
  \wHtot &=& (1+z)\frac{2 \bH'}{3 \bH} - 1 = \frac{2q_H-1}{3} = \frac{\Om + 4 \OQ - 1}{3} \ .
\eea

Correspondingly, the effective distance equation of state $w_D$ and the
effective total distance equation of state $\wDtot$ are defined to be those
of the spatially flat FRW model with the same distance $D_A(z)$
(and $\Omn$) as the backreaction model, \ie we replace $\bH$
by $\Hfit$ and $\Om$ by $\Omfit(z)\equiv\Omn (1+z)^3 (\bH_0/\Hfit(z))^2$
in \re{wH} and \re{wHtot}. We again use the redshift and the distance
calculated from the average expansion rate with \re{bz} and \re{DAH}.

\begin{figure}
%\hfill
\begin{minipage}[t]{7.7cm} 
\scalebox{1.0}{\includegraphics[angle=0, clip=true, trim=0cm 0cm 0cm 0cm, width=\textwidth]{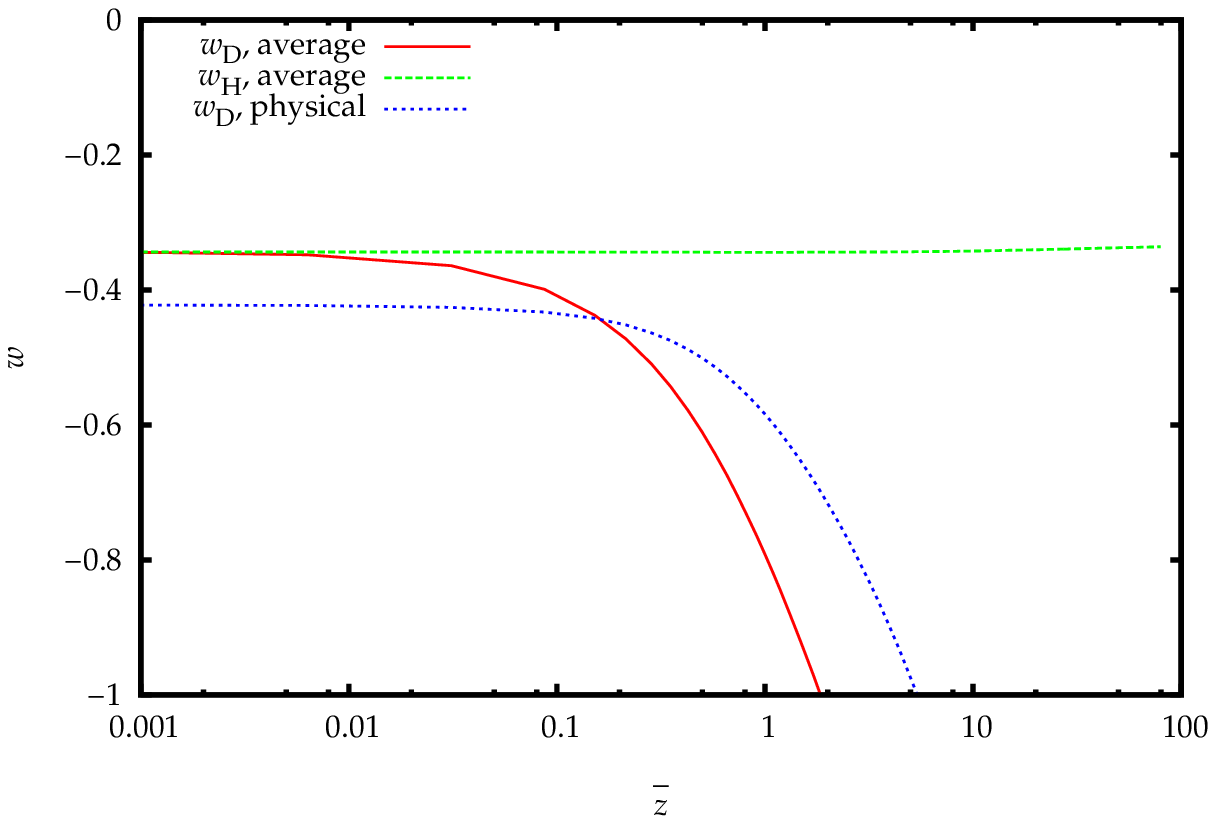}}
\begin{center} {\bf (a)} \end{center}
\end{minipage}
%\hfill
\begin{minipage}[t]{7.7cm}
\scalebox{1.0}{\includegraphics[angle=0, clip=true, trim=0cm 0cm 0cm 0cm, width=\textwidth]{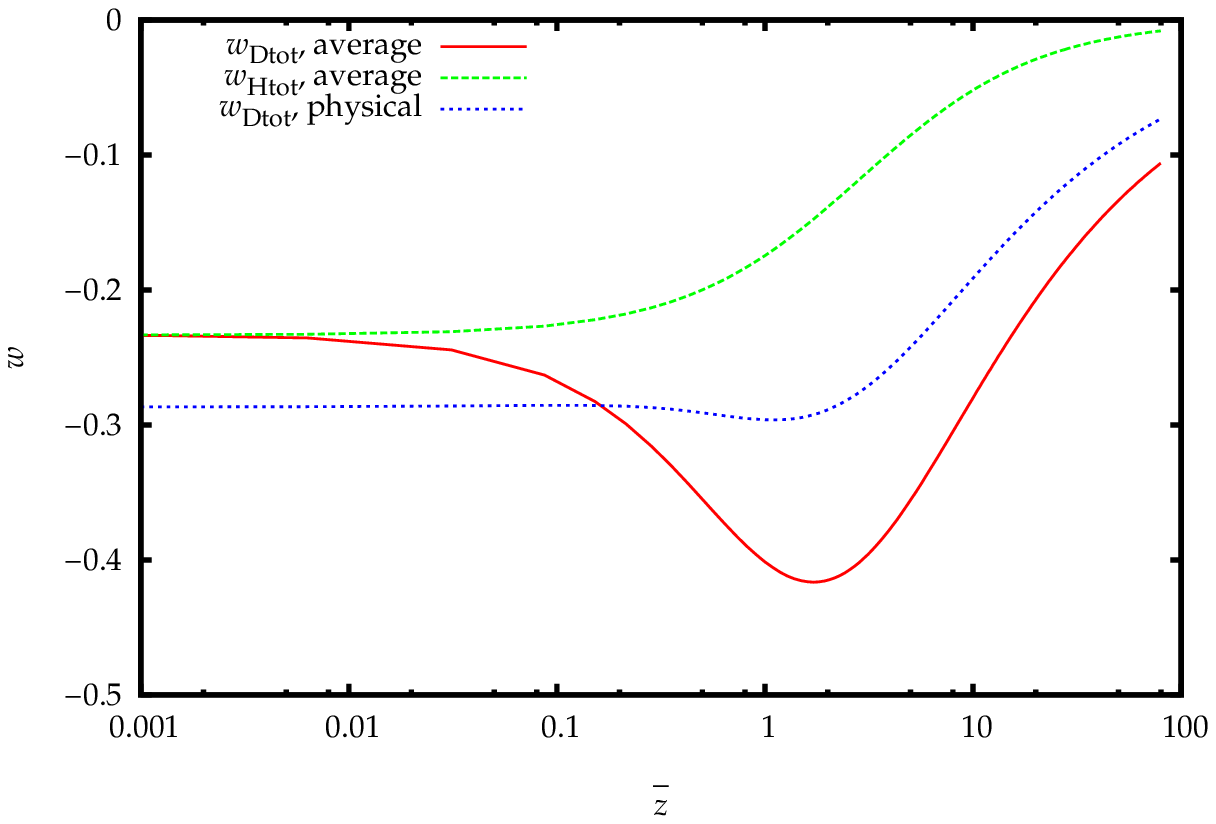}}
\begin{center} {\bf (b)} \end{center}
\end{minipage}
%\hfill
\caption{a) The effective equations of state $w_H$ and $w_D$.
b) The effective equations of state $\wHtot$ and $\wDtot$.}
\label{fig:w}
\end{figure}

We show the effective equations of state $w_H$ and $w_D$ in
\fig{fig:w}a, and $\wHtot$ and $\wDtot$ in \fig{fig:w}b, again
both for the distance and the redshift calculated from the average
expansion rate using \re{bz} and \re{DAH} and for the physical distance.
The evolution of $w_H$ is close to that of a dust FRW
model with an extra energy density component with
equation of state close to $-\frac{1}{3}$.
The equation of state $w_D$ is very different, but $\wDtot$
gives a better picture of the distance.
The behaviour of $\wDtot$ is qualitatively similar for both distances,
with a dip towards negative values. For the distance calculated from
the average expansion rate, $\wDtot$ goes below $-\frac{1}{3}$.
Correspondingly, the deceleration parameter
$q_D\equiv\frac{1}{2}(1+3\wDtot)$ shown in \fig{fig:exp}
is negative for a range of redshifts.
However, the deceleration parameter that corresponds to the physical
distance dips more moderately, and always remains positive.
So the model has neither apparent acceleration nor average
acceleration \cite{Rasanen:2011a, Buchert:2011}.
Figure \ref{fig:w}b demonstrates how the distance equation of
state $q_D$ is biased towards more negative values (in general,
towards $-1$) than $q_H$. As a result, average acceleration
may not be needed to explain the observations
\cite{Rasanen:2010b, Rasanen:2011a, Boehm:2013}. 
This biasing can also alleviate tension between having a negative
enough $q_H$ and a high enough $\bH t$.\footnote{A bound on $q_H$ in terms
of $\bH t$ was given in \cite{Rasanen:2011a}, but it may not hold if
rapidly collapsing regions are important, as in some toy models
\cite{Rasanen:2006a, Rasanen:2006b, Boehm:2013}.}

Such biasing is also present in the spatially curved FRW case.
The inferred equation of state can be strongly affected
if we try to determine the expansion rate of a FRW model assuming
that it is spatially flat when spatial curvature is in fact
non-zero \cite{Clarkson:2007a}.
For the distance calculated from the average expansion rate
in the present model, the biasing can be understood in the same way,
Even though the expansion rate is quite different
from the spatially flat background FRW model, it is close
to a negatively curved FRW model. This follows from the fact
that $\OQ$ is small (\ie $K$ is almost constant) at all times:
then the integrability condition \re{int} implies that
$\av{\sR}\propto a^{-2}$, just as in the FRW case.
Therefore, the difference between $w_H$ and $w_D$ can be understood
in terms of the FRW distance--expansion rate relation \re{DAFRW}:
if we try to interpret the distance of a FRW model with $K<0$ in terms
of a FRW model with $K=0$, the fitting model expansion rate will be
larger than the real expansion rate to reproduce the effect of
the $\sinh$ term.
However, this is not the case for the physical distance: the relation
between the average expansion rate and the physical distance cannot be
understood simply in terms of a FRW spatial curvature.
This is a general feature: if backreaction is significant, its effect
cannot be encapsulated in a ``global average FRW model'' that would
simultaneously account for both the expansion rate and the distance.
This is expected to be the case even for the distance calculated
from the average expansion rate, if the expansion first has extra deceleration
and then accelerates \cite{Rasanen:2006a, Rasanen:2006b, Boehm:2013}.
Such behaviour could be reproduced by having regions with $E<0$ in the holes.

\section{Discussion} \label{sec:disc}

\para{Integral formulation and series formulation.}

The reasoning behind relations \re{bz} and \re{DAH}
between the average expansion rate and redshift and distance
is that if we integrate the light propagation equations,
deviations around the average cancel because of statistical
homogeneity and isotropy \cite{Rasanen:2008b, Rasanen:2009b}.
Our results validate some aspects of this idea. In particular,
the average expansion rate gives a good description of the
redshift, though we have found that the contribution of the shear
is important, in contrast to the arguments of
\cite{Rasanen:2008b, Rasanen:2009b}.
If we neglect jumps in the Sachs equations due to the
surface layers, the angular diameter distance for $z\lesssim1$
is also accurately calculable from the average expansion rate.
The average expansion rate was also found to give a good description
in \cite{Bull:2012}, where the prescription was used that the time
that light propagates in a given different region is proportional to
the volume of the region, instead of considering light propagation
in a complete spacetime that is a solution of the Einstein
equation\footnote{The reason is that the time spent by a light ray in a
region scales linearly with the size of the region, and the probability of
a light ray entering a region scales quadratically.}.

In contrast to the integral formulation \re{DAH} used in the present
work and in \cite{Bull:2012}, the redshift and the distance have also
been considered in terms of a series expansion
\cite{Kristian:1966, Clarkson:2000, Clarkson:2011a}, which
seems to give different results. In particular, it has been suggested
that the distance can be quite different from the FRW case even when
the average expansion rate is close to FRW \cite{Clarkson:2011a}.
The series formulation also suggests large angular variation
in the distance. In \cite{Bull:2012} (see also \cite{Clifton:2013}),
the series formulation was found to disagree with the null
geodesic calculation, unlike the integral formulation.

One issue with the series expansion is that the redshift
along a null geodesic is in general not monotonic, so
the distance cannot be written as a function of the redshift
\cite{Rasanen:2008b}.
But even in models in which this is not the case, a series
expansion can be misleading.
If we approximate a function $f(z)$ between $z=0$
and $z=z_1$ with a Taylor series with terms up to $z^n$,
the remainder term is bounded by
$f^{(n+1)}_{\mathrm{max}} z^{n+1}/(n+1)!$, where
$f^{(n+1)}_{\mathrm{max}}$ is the maximum value of
$\rmd^{n+1} f/\rmd z^{\mathrm{n+1}}$ between $0$ and $z_1$.
Therefore, the series expansion is inaccurate for rapidly varying
functions. For non-analytic functions the series does not approximate
the function even for small $z$, even if it converges.
For example, consider the expansion rate as a function of redshift
for an observer located on the edge of a stabilised region such as a galaxy.
The expansion rate and its derivatives vanish
at the location of the observer. As the function
is not analytic, the Taylor series does not describe it.
More realistically, it could be said that physically the local expansion
rate is never exactly zero, and all functions can be approximated
by Taylor series. In this case the problem is that the derivatives are large.
The expansion rate changes rapidly by a factor of unity
between a galaxy and an unbound region, and other
quantities change even faster: the energy density varies
by orders of magnitude over distances which are tiny
on cosmological scales.
In the models considered in \cite{Bull:2012} and in the
present paper, variations are not so drastic, but there
are nevertheless strong variations on scales that are
small compared to the cosmological scale (in our case
the density varies by about a factor of 40 between the most
underdense and dense regions in a hole at the present time).

In the integral formulation, such rapid variations are
not important because they cancel out, leaving only
the average contribution. Averages of course depend on
the hypersurface on which they are taken, and the relevant
one for the line of sight integral is the hypersurface of
statistical homogeneity and isotropy
\cite{Rasanen:2006b, Rasanen:2008a, Rasanen:2008b, Rasanen:2011b, Rasanen:2009b, Geshnizjani, Rasanen:2004}.
In the present case, this is the same as the hypersurface of
constant proper time of the observers.

\para{Usefulness of the average expansion rate.}

It is perhaps surprising that the distance and the redshift
calculated from the average expansion rate give such
a good description of the physical distance (neglecting the
jumps due to the surface layers) and redshift.
After all, the average of the expansion rate is taken
on the three-dimensional hypersurface of constant proper
time, whereas cancellations in the deviations in the expansion
rate along the null geodesic happen in one dimension.
(It has even been questioned whether the average expansion
rate has any physical relevance at all \cite{Green}.)
However, in Euclidean space with statistically homogeneous and
isotropic (and static or slowly evolving) matter
distribution, one-dimensional averages converge
to three-dimensional averages, though this may require
distances much longer than the homogeneity
scale if the distribution is strongly clumped.
(In the case of curved spacetime, the measures are different
for a line integral and a volume integral, so the outcome
is less clear.)
For example, in the Millennium simulation, the matter
distribution is very filamentary, and the one-dimensional average
density seen by a typical light ray is about 20\% smaller
than the three-dimensional average even for distances of
1.5 Gpc \cite{Bolejko:2012b, Clarkson:2011c}.
However, deviation of the distance from the
background value is much smaller, less than 1\%
(see also \cite{Bolejko:2012a}).
Presumably this is related to the fact that the
distance depends on the density via a double integral.

In the limit where the matter distribution consists of
pointlike particles and the probability of a light
ray crossing any matter vanishes, one could expect
that the average expansion rate doesn't give a useful description at all.
However, the average expansion rate has been found to give
a useful description also in the discrete models studied in
\cite{Clifton, lattice}. (Though one might worry that the
geometrical optics approximation could break down \cite{Rasanen:2009b}.)
In our model, the holes occupy a large
fraction of space, so a light ray quickly samples
a representative distribution of the expansion rate:
a typical light ray spends approximately 50\% of the
travel time inside the holes.

It is sometimes claimed that light propagation in a clumpy
universe is on average identical to the FRW situation simply
because of flux conservation \cite{Weinberg:1976}.
However, the proof presented in \cite{Weinberg:1976} assumes that
the angular element is given by the FRW metric, which is the
question to be investigated \cite{Ellis:1998a}, and the 
LTB model has been used to provide an exact counterexample
\cite{Mustapha:1997} (see also \cite{Ellis:1998b}).
It has also been argued that the distance is close to FRW
as long as light travels through compensated over-
and underdensities, and the time spent by a light ray in a given
region is much smaller than the timescale for the evolution of
the gravitational potential of the region
\cite{DiDio:2011}. (This argument is tied to perturbation theory,
as the gravitational potential is a perturbative concept.)
However, our results, and those of \cite{Bull:2012}, support
the importance of the average expansion rate, not only
the average density as in the Dyer-Roeder approximation
\cite{Rasanen:2008b, Rasanen:2009b} (see also
\cite{Bull:2012, Bolejko:2012a, Clarkson:2011c, Ehlers:1986}).
It is the integrated effect over all regions that is relevant,
not the time spent by the light in a given region.
However, in order for the average expansion rate to give a good
description, it is necessary that the distribution of structures
does not change significantly during the time that it takes for a
light ray to travel a spatial distance equal to the homogeneity scale.

\para{Relation to Newtonian gravity.}

In the spherically symmetric subcase, the result that backreaction
is small is related to the fact that in Newtonian gravity, the average
expansion rate of a spherical system is identical to the FRW case
\cite{Buchert:1999b}.
This result is well known under the name of spherical
collapse model. It is sometimes claimed that the result would
extend to general relativity (at times this is even ascribed to
Birkhoff's theorem; see section 2.5 of \cite{Buchert:2011}).
While this is not true in general, the spatially flat LTB model
with $E=0$ is close to Newtonian gravity \cite{Szekeres:1999}.
The structure of the general quasispherical Szekeres model
is similar to that of the LTB model, as the metric depends on
time only via $R(t,r)$, and for small, long-lived regular holes
$E$ is small.
From the geometrical point of view, the appearance of general
relativistic degrees of freedom in the Szekeres model is constrained
by the fact that the magnetic part of the Weyl tensor is zero.
This is always true in Newtonian theory, but in general relativity
dust models with vanishing magnetic Weyl tensor are a very limited
class of solutions
\cite{Ellis:1971, Ellis:1994, Kofman:1995, Matarrese:1995, Ehlers:2009, Rasanen:2010a, Rasanen:2011b, Bertello:2012}.
It is conjectured that the Szekeres model is the only such
inhomogeneous and irrotational solution \cite{silent} (a proof was
claimed in \cite{Apostolopoulos:2007}); whether the class of such
solutions with rotation is non-empty is not known.
The magnetic part of the Weyl tensor is non-zero in realistic
solutions: for example, it is essential in gravitational collapse
\cite{Kofman:1995}. Its importance for the backreaction
of realistic structures is not clear.

\section{Conclusion} \label{sec:conc}

\para{Summary and outlook.}

We have proven that the average expansion rate in a Swiss Cheese
dust model with Szekeres holes is close to the FRW model if the holes
are small and long-lived, there are no singularities, the
center of the hole is regular and the metric function
$R$ is monotonic in the coordinate radius.
By violating the last assumption, we have then built
the first exact statistically homogeneous and isotropic solution
in which inhomogeneity has a significant effect on the average expansion rate,
\ie backreaction is large. The expansion rate is close to the FRW
case at early times, but increases relative to the background at late times.
However, in the model, violating the monotonicity of $R$ requires
surface layers.

We have studied the relation of the average expansion rate to
the angular diameter distance and the redshift.
The surface layers lead to large jumps in the area expansion rate
of light bundles and sharp turning of light rays.
We therefore consider a modified version of the distance in which
we neglect the jumps due to the surface layers.
For the redshift, we consider both the case when the light rays
turn sharply as well as a modification in which we take them to
follow straight paths as defined by the background metric.
We find that the modified angular diameter distance is significantly
different from the background, but it is fairly well described by the
distance calculated from the average expansion rate up to redshifts of
order unity, though not for larger redshifts.
The redshift also differs markedly from the background value,
and it is well described by the average expansion rate, with
statistical fluctuations of less than 10\%  for the unmodified rays,
and less than 5\% for the straight rays.
In contrast to expectation, the expansion rate along the
light ray is not the same as the spatial average, though the
difference cancels with the contribution of shear along the light ray.
This situation is the same both for unmodified and straight rays.

The results show the usefulness of the spatially averaged
expansion rate in describing light propagation.
However, as the unrealistic surface layers have a large effect
on light propagation, it would be interesting to consider
models without them.
One simple and physically motivated way to extend the present
study and bypass our Szekeres Swiss Cheese theorem without surface
layers would be to resolve unphysical shell crossing and collapse
singularities either by introducing pressure \cite{Bolejko:2008c}
or by applying some other prescription to include interactions
that prevent singularities in real structure formation \cite{collapse}.

Another issue related to the surface layers is that although
the local expansion rate, shear and density asymptotically
approach FRW values at early times, this is not the case for
the metric nor the spatial curvature.
It would be interesting to construct exact solutions in which the
metric is perturbatively close to FRW at early times, but
backreaction is nevertheless large at late times.

\acknowledgments

SR thanks Krzysztof Bolejko and Roberto Sussman for correspondence.
SJS was supported by the Polish Ministry of Science and Higher Education
(the Iuventus Plus grant no.\ IP2011055071) and would like to thank the
University of Helsinki for hospitality.

\end{document}